\documentclass[preprintnumbers,10pt,amsmath,amssymb]{revtex4}
\usepackage{url} 
\usepackage{graphicx}
\usepackage{amsthm}
\usepackage{enumerate}

\newtheorem{theorem}{Theorem}[section]

\newtheorem{lemma}[theorem]{Lemma}
\newtheorem{corollary}[theorem]{Corollary}
\newtheorem{definition}[theorem]{Definition}
\newtheorem{remark}[theorem]{Remark}
\newtheorem{example}[theorem]{Example}

\newcommand*{\myspan}{\textsf{span}}
\newcommand*{\unitary}{\mathsf{g}} 
\newcommand*{\gel}{\mathsf{g}} 
\newcommand*{\specialunitarygroup}[1]{{\cal SU}(#1)} 
 
\newcommand*{\weights}[1]{\cW_{#1}}
\newcommand*{\weight}[1]{{\mathbf #1}}
\newcommand*{\w}{\weight{w}}
\newcommand*{\weightspace}[2]{#1^{#2}}
\newcommand*{\heighthigh}[2]{ht^\downarrow_{#1}(#2)}
\newcommand*{\heightlow}[2]{ht^\uparrow_{#1}(#2)}
\newcommand*{\weightshigh}[2]{\cW_{#1}^{\downarrow r}}
\newcommand*{\weightslow}[2]{\cW_{#1}^{\uparrow r}}
\newcommand*{\weightr}[2]{\cW_{#1}^{#2}}

\newcommand*{\textfrac}[2]{{#1}/{#2}}
\newcommand*{\id}{\mathsf{1}} 
\newcommand*{\complex}{\bbC}
\newcommand*{\heis}{\mathsf{H}}

\newcommand*{\cA}{\mathcal{A}}
\newcommand*{\cB}{\mathcal{B}}
\newcommand*{\cC}{\mathcal{C}}

\newcommand*{\cG}{\mathcal{G}}
\newcommand*{\cH}{\mathcal{H}}

\newcommand*{\cK}{\mathcal{K}}

\newcommand*{\cN}{\mathcal{N}}

\newcommand*{\cR}{\mathcal{R}}

\newcommand*{\cU}{\mathcal{U}}

\newcommand*{\cV}{\mathcal{V}}
\newcommand*{\cW}{\mathcal{W}}
\newcommand*{\cX}{\mathcal{X}}

\newcommand*{\tr}{\mathsf{tr}}
\newcommand*{\ket}[1]{|#1\rangle}
\newcommand*{\bra}[1]{\langle #1|}
\newcommand*{\proj}[1]{\ket{#1}\bra{#1}}
\newcommand*{\Sym}[2]{\mathsf{Sym}^{#2}(#1)}

\newcommand{\be}{\begin{equation}}
\newcommand{\ee}{\end{equation}}
\newcommand{\bea}{\begin{eqnarray}}
\newcommand{\eea}{\end{eqnarray}}
\newcommand{\bestar}{\begin{equation*}}
\newcommand{\eestar}{\end{equation*}}
\newcommand{\beastar}{\begin{eqnarray*}}
\newcommand{\eeastar}{\end{eqnarray*}}
\newcommand*{\spr}[2]{\langle #1|#2\rangle}
\def\complex{\mathbb{C}}
\newcommand*{\dimU}[2]{f_{#1,#2}}
\newcommand*{\liealgebra}{\mathfrak{g}}

\begin{document}
\title{A most compendious and facile quantum de Finetti theorem}

\author{Robert \surname{K\"onig}}
\email[]{r.t.koenig@damtp.cam.ac.uk} \affiliation{Centre for
Quantum Computation, DAMTP,
             University of Cambridge,
             Cambridge CB3 0WA, UK}

\author{Graeme \surname{Mitchison}}
\email[]{g.j.mitchison@damtp.cam.ac.uk} \affiliation{Centre for
Quantum Computation, DAMTP,
             University of Cambridge,
             Cambridge CB3 0WA, UK}

\begin{abstract}
In its most basic form, the finite quantum de Finetti theorem states
that the reduced $k$-partite density operator of an $n$-partite
symmetric state can be approximated by a convex combination of
$k$-fold product states. Variations of this result include Renner's
"exponential" approximation by "almost-product" states, a theorem
which deals with certain triples of representations of the unitary
group, and D'Cruz et al.'s result for infinite-dimensional systems. We
show how these theorems follow from a single, general de Finetti
theorem for representations of symmetry groups, each instance
corresponding to a particular choice of symmetry group and
representation of that group. This gives some insight into the nature
of the set of approximating states, and leads to some new results,
including an exponential theorem for infinite-dimensional systems.
\end{abstract}

\maketitle

\section{Introduction}
Edmund Halley, of the comet, used the opening words of our title to
describe a method for calculating logarithms, making it clear that his
principal claim was to an increased ease and scope~\cite{halley65}.
This is our claim too: our main theorem gathers together the currently
available examples of quantum de Finetti theorems and gives them a
larger setting and what we hope is an illuminating and accessible
proof.

Our theorem is of the general type first proved in~\cite{Ren05} and
recently reviewed in~\cite{Ren07}. Recall that the standard de Finetti
theorem says that the state obtained by tracing out $n-k$ parts from a
symmetric $n$-partite state can be approximated by a convex sum of
product states of the form $\sigma^{\otimes
k}$~\cite{KoeRen05,chrkoemire06}, the error in the approximation being
of size $O(k/n)$ for a fixed dimension of the individual
subsystems~\cite{chrkoemire06}. This theorem has many
uses~\cite{holeop06,chiribella06,fuchsschacksecond,Hudson81,fannes,raggiowerner,bruncavesschack,dohertyetal,audenart,terhaldoherty,BaeAcin06},
but for some purposes one needs the type of theorem mentioned above,
where the class of approximating states is broadened and the
convergence is much more rapid~\cite{Ren05,holeop06}. The
approximating states in question are {\em almost-product states}~\cite{Ren05}, that are sums of terms of the form $\ket{\psi_1}
\otimes \ket{\psi_2} \otimes \cdots \otimes \ket{\psi_k}$, where all
but a few of the $\ket{\psi_i}$ are identical. The convergence now has
an exponential character, with an error of order $e^{-\gamma n}$ for
some $\gamma>0$ (where $\gamma$ depends on $k/n$ and the fraction of
$\ket{\psi_i}$ that differ; see Corollary~\ref{cor:almostproduct}). We
shall therefore refer to this as an {\em exponential} theorem; it
implies that, with almost product states, far fewer subsystems have to
be traced out to get a close approximation. For many physical
questions, the fact that a few subsystems differ from the rest has
only a small effect; for instance, most thermodynamic measures will
register little difference between product states and almost-product
states~\cite{Ren07}.

The setting for our theorem is representation theory, and the class of
approximating states is determined by some subspace $\cX$ of the
representation. One case of interest is where there is a weight space
structure, and where $\cX$ consists of high weight vectors. Such vectors
are closely related to almost-product states. To see how this
works, consider the representation of $\specialunitarygroup{d}$ on the
symmetric subspace $\Sym{\complex^d}{k}$ of $(\complex^d)^{\otimes
k}$. The highest weight vector is $\ket{1}^{\otimes k}$, and any
product state $\ket{\psi}^{\otimes k}$ can be obtained by applying
some element $\gel$ of $\specialunitarygroup{d}$ to this vector. Thus
a product state is a rotated highest weight vector. There is an
ordering of weights, the next highest weight vector being
$\left(\ket{211 \cdots 11}+\ket{121 \cdots 11} + \cdots +\ket{111
\cdots 12}\right)/\sqrt{k}$, and a symmetric almost-product state with
all but one of its factors identical can be obtained by applying some
$\gel\in\specialunitarygroup{d}$ to this vector.

The original exponential theorem in~\cite{Ren05} can be obtained from
the symmetric representation $\Sym{\complex^d}{k}$ in this way.  The
theorem for unitary representations in~\cite{chrkoemire06} is another
corollary, in a more general form that allows all types of
representations and extends the class of approximating states. A
further corollary shows that, for certain representations, the de
Finetti approximation becomes exact. Finally, using representations of
the Heisenberg group, we obtain the de Finetti theorem recently proved
for coherent states of infinite-dimensional
systems~\cite{CruzOsborneSchack06} together with a new exponential version of it.

Having laid claim to an all-embracing theorem, we must acknowledge
that there are some special de Finetti results that do not lie within
our conspectus. One is the theorem for unitarily-invariant, symmetric
mixed states proved in~\cite{chrkoemire06}. Though the theorem in the
present paper applies to pure states, it can be extended by symmetric
purification to mixed states. However, this does not yield the special
form of the approximation given in~\cite{chrkoemire06}. Nor does our
theorem apply to the unitarily-invariant states considered
in~\cite{Mit07}; but here unitary invariance takes the place of
symmetry, and the result is of a very different character from all
other de Finetti theorems.

\section{Background on representations \label{sec:background}}
We first recall some basic facts about
representations (for more details, see e.g.,~\cite{perelomov86,humphreys87,CarterSegalMacDonald95,fulharr99,goodmanwallach03,knapp05}). Section~\ref{sec:gl} deals with the group of special
unitary matrices, whereas Section~\ref{sec:heisenbergweylintro}
discusses the Heisenberg group. Throughout, we will only consider
 representations on complex Hilbert spaces $\cH$ which are unitary,
 i.e.  the group elements act as unitary operators on $\cH$.

\subsection{The special unitary group $\specialunitarygroup{d}$ and its representations\label{sec:gl}}

Let $\specialunitarygroup{d}$ denote the group of unitary matrices on
$\complex^d$ with determinant~$1$, and let $\ket{1}, \ldots \ket{d}$
be a chosen basis for~$\complex^d$. Let $H$ denote the subgroup of
diagonal matrices with respect to this basis. Let $\w=(w_1, \ldots,
w_d)$ be a vector of integers with $w_i \ge 0$, and let $\cR$ be a
representation of $\specialunitarygroup{d}$. Then a {\em weight
vector} of $\cR$ with weight $\w$ is a vector $\ket{\w} \in \cR$
satisfying $h\ket{\w}=\prod h_i^{w_i}\ket{\w}$, where $h$ is a
diagonal matrix in $H$ with diagonal entries $h_1, \ldots, h_d$. For
instance, the weight $(n,0, \ldots, 0)$ corresponds to the weight
vector $\ket{1}^{\otimes n}$ in the symmetric representation
$\Sym{\complex^d}{n}\subset(\mathbb{C}^d)^{\otimes n}$.  In general,
there may be several weight vectors for a given weight. In the sequel,
we will write $\ket{\w}$ whenever we refer to a normalised vector of
weight $\w$.  These should not be confused with basis vectors
$\ket{i}$, which are labelled by $i\in\{1,\ldots,d\}$. We will also
distinguish different vectors of the same weight $\w$ by superscripts
$\ket{\w}^{i}$.

A special part is played by irreducible representations of
$\specialunitarygroup{d}$, i.e. representations that cannot be written
as a direct sum of two subrepresentations. Because of the
correspondence between Lie groups and their Lie algebras (see
Section~\ref{sec:semi-simple}), the study of these representations is
equivalent to the study of irreducible representations of the (real)
Lie algebra $\mathfrak{su}_d$ consisting of anti-hermitian traceless
matrices. Irreducible representations of $\mathfrak{su}_d$ are in turn
in one-to-one-correspondence with irreducible representations of its
complexification $\mathfrak{sl}_d(\complex)$, which consists of all
traceless matrices.  Thus it suffices to consider the Lie
algebra~$\mathfrak{sl}_d(\mathbb{C})$ instead of the Lie group
$\specialunitarygroup{d}$, which often turns out to be simpler.

An irreducible representation of $\specialunitarygroup{d}$ (or
equivalently $\mathfrak{sl}_d(\mathbb{C})$) has a unique highest weight (we
will discuss the ordering of weights below), and two irreducible
representations are equivalent if and only if they have the same
highest weight. The weight space corresponding to the highest weight
is one-dimensional. We write $\cR_\lambda$ for the irreducible
representation of $\specialunitarygroup{d}$ with highest weight
$\lambda$ (Greek letters are conventionally used), and the normalised
highest weight vector of $\cR_\lambda$ will be denoted by
$\ket{\weight{\lambda}}$. The highest weights have the property that
$\lambda_1 \ge \lambda_{2}\geq \cdots \geq \lambda_d=0$. We will often
omit $\lambda_i$ that are zero, for instance writing the highest
weight of the symmetric representation $\Sym{\complex^d}{n}$ as $(n)$
rather than $(n,\underbrace{0,\ldots,0}_{d-1})$.  We will also
sometimes write $\cR_{\lambda}$ for an irreducible representations
with highest weight $\lambda=(\lambda_1,\ldots,\lambda_d)$ where
$\lambda_d\neq 0$; note that this is equivalent to an irreducible
representation $\cR_{\lambda'}$ where $\lambda'$ is obtained from
$\lambda$ by subtracting $\lambda_d\cdot (1,\ldots,1)$.

There is a special set of weights, of the form 
$\alpha_i=(0,\ldots,0,1,-1,0,\ldots,0)$ where the $i$-th entry is $1$.
These are a set of so-called {\em simple roots} of the
Lie algebra $\mathfrak{sl}_d(\complex)$.  Any weight occurring in the irreducible
representation $\cR_\lambda$ can be obtained by subtracting integer
combinations of the $\alpha_i$ from the highest weight vector. This
allows us to define a notion of the {\em height} of a weight
$\weight{w}$ in the irreducible representation $\cR_\lambda$ as
$ht^{\downarrow}_\lambda(\weight{w}):=\max_i |n_i|$, where
$\weight{w}=\lambda-\sum n_i\alpha_i$.  More generally, we will 
extend this definition to the space $\mathbb{R}^n$ of all $n$-tuples
$\weight{w}=(w_1,\ldots,w_n)$ (not necessarily weights); this is well-defined because
the $n_i$ for a given $\weight{w}$ are uniquely determined, as the simple roots are a basis of $\mathbb{R}^n$. There is
an analogous formulation using the lowest weight vector $\lambda_*$ of
$\cR_\lambda$. Any weight in $\cR_\lambda$ can be obtained by adding
integer combinations of the $\alpha_i$ to $\lambda_*$, so we can
define $ht^{\uparrow}_\lambda(\weight{w}):=\max_i |m_i|$, where
$\weight{w}=\lambda_*+\sum m_i\alpha_i$.

There is a partial ordering on the weights, with $\w \preceq \w^\prime$ if 
the difference $\w^\prime-\w$ is a combination of the simple roots $\alpha_i$ with nonnegative coefficients, which is equivalent to the condition $\sum_{i=1} ^\ell w_i\leq \sum_{i=1} ^\ell w^\prime_i$ for all $i=1,\ldots,\ell$. The set
of weights in a representation $\cR$ will be denoted by
$\weights{\cR}$. Given some subset $\cW \subset \weights{\cR}$, the
subspace of $\cR$ that it generates will be denoted by
$\weightspace{\cR}{\cW}$. For a finite-dimensional representation, the weight vectors generate the
whole representation, so we have $\weightspace{\cR}{\cW_\cR}=\cR$.

Given two representations $\cA$ and $\cB$, not necessarily
irreducible, we can define the tensor product representation $\cA
\otimes \cB$ on which any $\unitary \in \specialunitarygroup{d}$ acts by the
tensor product of its actions on $\cA$ and $\cB$. Since the weight
vectors of $\cA$ and $\cB$ span each representation, a weight vector
$\ket{\w}_{\cA \otimes \cB}$ of the tensor product can be written as a
sum of products of weight vectors as
\be \label{weight-expansion}
\ket{\w}_{\cA \otimes \cB}=\sum_{\w_\cA,\w_\cB,i,j} \gamma_{ij}^{\w_\cA\w_\cB} \ket{\w_\cA}^{i}_\cA
\otimes \ket{\w_\cB}^{j}_\cB\ ,
\ee
and it follows from the definition that the weights occuring with
non-zero coefficent $\gamma_{ij}^{\w_\cA\w_\cB}$ satisfy
\be \label{tensor-sum}
\w_\cA + \w_\cB=\w\ .
\ee

The tensor product of two irreducible representations $\cR_\mu$ and $\cR_\nu$ is in general
reducible and decomposes as
\be \label{CGexpansion}
\cR_\mu \otimes \cR_\nu \cong \bigoplus_{\lambda} c_{\mu \nu}^\lambda
\cR_\lambda,
\ee
where the multiplicities $c_{\mu \nu}^\lambda$ are the
\emph{Littlewood-Richardson coefficients}. The subrepresentations on
the right-hand side include $\cR_{\mu+\nu}$, which occurs with
multiplicity 1.

For $d=2$, the multiplicities $c_{\mu\nu} ^\lambda$ are all~$1$ and furthermore there is a
unique weight vector for each weight. Thus equation~\eqref{weight-expansion} can be written more simply as
\be \label{twoD}
\ket{\w}_\lambda=\sum_{\w^\prime} a_{\w,\w^\prime} \ket{\w^\prime}_\mu
\otimes \ket{\w-\w^\prime}_\nu\ ,
\ee
and the coefficients $a_{\w,\w^\prime}$ define the inclusion map
$\cR_\lambda \to \cR_\mu \otimes \cR_\nu$ in
(\ref{CGexpansion}). These are the {\em Clebsch-Gordan coefficients},
and there seems to be no consensus on notation for them. Here we write
$a_{\w,\w^\prime}$ as $\langle j_1 m_1 j_2 m_2 | j m \rangle$, where
$j_1=(\mu_1-\mu_2)/2$, $j_2=(\nu_1-\nu_2)/2$,
$j=(\lambda_1-\lambda_2)/2$, $m=(w_1-w_2)/2$,
$m_1=(w_1^\prime-w_2^\prime)/2$, and $m_2=m-m_1$.  In other words
$\{\ket{jm}\}_{m=-j}^j$ is a basis of weight vectors of $\cR_{(2j)}$,
where $\ket{jm}$ has weight $(m,-m)$. Thus~\eqref{twoD} can
equivalently be written as
\begin{align}
\ket{jm}=\sum_{m_1+m_2=m} \spr{j_1m_1j_2m_2}{jm}\ket{j_1m_1}\otimes\ket{j_2m_2}\ .
\end{align}
Here $j_1,j_2\in\{0,\textfrac{1}{2},1,\ldots,\}$, and
$\cR_{(2j)}\subset\cR_{(2j_1)}\otimes\cR_{(2j_2)}$ with
multiplicity~$1$ for all $j\in\{|j_1-j_2|,|j_1-j_2|+\textfrac{1}{2},\ldots,j_1+j_2\}$.

\subsection{Representations of the Heisenberg group\label{sec:heisenbergweylintro}}
The Heisenberg group $\heis$ is
$\mathbb{C}\times\mathbb{R}$ with multiplication
\[
(\alpha;t)(\beta;t')=(\alpha+\beta;t+t'+\Im(\alpha\bar{\beta}))\ ,\]
where $\Im(\gamma)$ denotes the imaginary part of
$\gamma\in\mathbb{C}$.  The centre of $\heis$ is $Z=\{(0;t)\ |\
t\in\mathbb{R}\}$, and the quotient group $\heis/Z$ is isomorphic to
the abelian group $\mathbb{R}^2$. Elements of $\heis/Z$ will be
denoted simply by a complex number $\alpha\in\mathbb{C}$.  The
irreducible infinite-dimensional unitary representations of $\heis$
are determined by a real number $\lambda\neq 0$. The
action of Z in the irreducible representation
$\cH_\lambda$ is then given by
\begin{align}\label{eq:lambdairrep}
(0;t)\ket{\psi}=e^{i\lambda t}\ket{\psi}\qquad\textrm{for every
}\ket{\psi}\in\cH_\lambda\ .
\end{align}

The representation $\cH_1$ can be obtained from an {\em annihilation
 operator} $a$ and its conjugate {\em creation operator} $a^\dagger$
 acting on some (infinite-dimensional) Hilbert space $\cH$ satisfying
 the canonical commutation relation $[a,a^\dagger]=\id$ and a ``vacuum
 state'' $\ket{0}\in\mathbb{\cH}$ satisfying $a\ket{0}=0$. The space
 $\cH_1$ is then spanned by the orthonormal vectors
 $\{\ket{n}\}_{n\in\mathbb{N}_0}$, where
 $\ket{n}:=\frac{(a^\dagger)^n}{\sqrt{n!}}\ket{0}$ for every
 nonnegative integer $n$. To define the action of $\heis$ on this space,
 we introduce the {\em displacement operators}
\begin{align}\label{eq:displacementoperatordef}
D(\alpha)=\exp(\alpha a^\dagger-\bar{\alpha}a)=e^{-\textfrac{|\alpha|^2}{2}}\exp(\alpha a^\dagger)\exp(-\bar{\alpha}a)
\end{align}
for $\alpha\in\mathbb{C}$. (We will sometimes write $D_a(\alpha)$ to
clarify what the operators $a$ and $a^\dagger$ in this definition
are.) It is straightforward to verify that these operators satisfy the
relation
\begin{align}\label{eq:weylheisenbergsecond}
D(\alpha)D(\beta)=e^{i\Im(\alpha\bar{\beta})}D(\alpha+\beta)\ .
\end{align}
This identity implies that an  action of $\heis$ on $\cH_1$ is defined by
\begin{align}\label{eq:weylheisenbergfirst}
(\alpha;t)\ket{\psi}:=e^{i t} D(\alpha)\ket{\psi} \qquad\textrm{for
  all }\ket{\psi}\in\cH_1\ .
\end{align}
This completes the description
of $\cH_1$. 
Note that the creation and annihilation operators satisfy 
\begin{align}
a^\dagger
\ket{n}=\sqrt{n+1}\ket{n+1}\qquad\textrm{ and }\qquad
a\ket{n}=\sqrt{n}\ket{n-1}\ ,\label{eq:creationannihilationaction}
\end{align} 
and the state $\ket{n}$ is an eigenstate of the number operator
$a^\dagger a$ with eigenvalue $n$. The subspace spanned by a set $\cN$
of number states plays the same role as the subspaces spanned by
weights $\cW$ in the case of representations of
$\specialunitarygroup{d}$. The number state $\ket{0}$ is analogous to
a lowest weight vector, and the corresponding displaced states
$D(\alpha)\ket{0}=e^{-\textfrac{|\alpha|^2}{2}}\sum_{n=0} ^\infty
\frac{\alpha^n}{\sqrt{n!}}\ket{n}$ for $\alpha\in\mathbb{C}$ are
commonly called {\em coherent states}.

To give an explicit construction of $\cH_\lambda$ for $\lambda\neq 0$,
we choose the space $\cH_\lambda$ as the span of $\{\ket{n}\ |\
n\in\mathbb{N}_0\}$ and define
\begin{align}\label{eq:hlambdaaction}
(\alpha;t)\ket{\psi}:=e^{i\lambda
  t}D(\sqrt{\lambda}\alpha)\ket{\psi}\qquad\textrm{for all
  }\ket{\psi}\in\cH_\lambda\ ,
\end{align}
where the states $\ket{n}$ and the operators $D(\alpha)$ are defined
as before. Again using~\eqref{eq:weylheisenbergsecond}, it is
straightforward to check that this defines a representation, and it is
irreducible as a consequence of the fact that $\cH_1$ is irreducible.

A few subtleties arise when integrating over the group. The quotient
group $\heis/Z$ is a {\em unimodular group}, meaning that there is a left-
and right-invariant Haar measure $\mu$ which assigns finite mass
$\mu(K)$ to every compact set $K\subset \heis/Z$. This Haar measure is given
by the standard Lebesque measure on $\mathbb{C}^2$; unlike the case of
a compact group $\cK$, this can not be normalised so that
$\mu(\cK)=1$. Thus the Haar measure on a unimodular group is only
fixed up to a constant. For reasons that will become clearer below, we
will choose the measure on $\heis/Z$ as
$d\mu(\alpha)=\frac{1}{\pi}d\Re(\alpha)d\Im(\alpha)$ on $\heis/Z$.

A version of Schur's lemma which applies to general unimodular groups
$\cG$ involves the notion of a {\em square-integrable representation}
on a Hilbert space $\cH$. Such a representation has the property that
\[
\int_{\cG} |\spr{\psi}{\gel|\varphi}|^2d\mu(\gel)<\infty \qquad
\textrm{for all }\ket{\psi},\ket{\varphi}\in\cH\ .
\]
It is known~\cite[p.~439,~Proposition 29]{gaal73} that if $\cG$ is
unimodular and the representation $\cH$ irreducible and
square-integrable, then there is a constant $d_\cH$ such that
\begin{align}\label{eq:squareintegrableprop}
\int_{\cG} \spr{\beta}{\gel^\dagger|\alpha}
\spr{\gamma}{\gel|\delta}d\mu(\gel)=\frac{\spr{\gamma}{\alpha}\spr{\beta}{\delta}}{d_\cH}\ 
\end{align}
for all
$\ket{\alpha},\ket{\beta},\ket{\gamma},\ket{\delta}\in\cH$. The
quantity $d_\cH$ is called the {\em formal degree} of $\cH$. Note that
it depends on the intitial choice of the Haar measure $\mu$. For a
finite-dimensional representation $\cH$ and a compact group $\cG$,
the formal degree $d_\cH$ is equal to the dimension of $\cH$ if the
Haar measure is normalised so that $\int_\cG d\mu(\gel)=1$.

Note that the representation $\cH_\lambda$ of the Heisenberg group~$\heis$ defined by~\eqref{eq:hlambdaaction} gives an irreducible
representation of the unimodular quotient group $\heis/Z$. For the chosen
Haar measure, the formal degree of this representation can easily be
computed using the fact that $|\langle
0|D(\alpha)|0\rangle|^2=e^{-|\alpha|^2}$. One finds
$d_{\cH_\lambda}=\lambda$. This is all we need for the Heisenberg
group; we refer the reader to the literature
(e.g.,~\cite{perelomov86}) for more details.

\section{The main theorem\label{sec:approximationwstates}}
We consider irreducible representations $\cA$, $\cB$ and $\cC$ of a
general unimodular group $\cG$ satisfying
$\cC\subset\cA\otimes\cB$. The trace $\tr_{\cB}\proj{\Psi}$ of a state
$\ket{\Psi} \in \cC$ is now well-defined, and our aim is to show that
this trace can be approximated by a convex sum of a special class of
states given by:
\begin{definition}
Given a subset $\cX$ of $\cA$, an $\cX$-state is a state of the form
$\gel\ket{\varphi}$, where $\gel \in \cG$ and $\ket{\varphi} \in \cX$.
\end{definition}
Now suppose we have a triple of representations $\cA$, $\cB$ and
$\cC\subset\cA\otimes\cB$ with $\cB$ and $\cC$ square-integrable (see
Section \ref{sec:heisenbergweylintro}).
\begin{definition}\label{def:delta}
Let $\ket{\psi}\in\cB$ be arbitary, and 
let 
\[
\delta_{\ket{\psi}}(\cX)=\frac{d_\cB}{d_\cC}\tr[P_{\cC}(P_\cX\otimes\proj{\psi})]\ ,
\]
where $P_{\cC}$ and $P_\cX$ are the projectors onto the subspaces
$\cC\subset\cA\otimes\cB$ and $\cX \subset\cA$, and
$d_\cB$, $d_\cC$ are the formal degrees of $\cB$ and $\cC$,
respectively. Finally, let
\[
\delta(\cX)=\sup_{\ket{\psi}\in\cB} \delta_{\ket{\psi}}(\cX),
\]
where the maximisation is over all normalised pure states on $\cB$.
\end{definition}
We will discuss two basic properties of this definition below
(Lemma~\ref{lem:deltaproperties}). The following theorem, which is our
main result, shows that $\delta(\cX)$ is a useful
measure for the error in the de Finetti approximation.
\begin{theorem}[Approximation by $\cX$-states]\label{thm:highweight}
Let $\cA$, $\cB$ and $\cC\subset\cA\otimes\cB$ be irreducible
representations of a unimodular group~$\cG$, where $\cB$ and $\cC$ are
square-integrable. Let $\cX$ be a finite-dimensional
subspace of $\cA$.  Then $\tr_{\cB}\proj{\Psi}$ for every
$\ket{\Psi}\in\cC$ can be approximated by a convex combination of
$\cX$-states with error $2\sqrt{1-\delta(\cX)}$. That is, there is a
probability measure $m$ on $\cG$ and states $\ket{\chi_{(\gel)}}\in
\gel \cX$ for $\gel\in\cG$ such that
\begin{align}\label{eq:highweightapproximationerr}
\bigl\|\tr_{\cB}\proj{\Psi}-\int \proj{\chi_{(\gel)}}dm(\gel)\bigr\|\leq
2\sqrt{1-\delta(\cX)}\ .
\end{align}
In this expression, the {\em trace norm} is defined as
$\|A\|=\frac{1}{2}\tr(\sqrt{A^\dagger A})$ for any operator $A$.
\end{theorem}

\begin{remark}\label{rem:improvement}
In certain cases, it is possible to improve the
bound~\eqref{eq:highweightapproximationerr} by adapting the proof
technique introduced in~\cite{chrkoemire06} and applied
in~\cite{chiribella06,CruzOsborneSchack06,Ren07}. For example, if the
representation $\cC$ appears with multiplicity $1$ in the tensor
product $\cA\otimes\cB$, then the
rhs.~of~\eqref{eq:highweightapproximationerr} can be replaced by
$2(1-\delta(\cX))$. Also, if $\cX$ has an orthonormal basis
$\{\ket{\varphi_1},\ldots,\ket{\varphi_\ell}\}$ and there is a vector
$\ket{\psi}\in\cB$ such that $\ket{\varphi_i}\otimes\ket{\psi}\in\cC$
for all $i$, then we can replace the
rhs. of~\eqref{eq:highweightapproximationerr} by
$2(1-\delta_{\ket{\psi}}(\cX))$.  This improvement is made possible by
the fact that twirling the operator $P_\cX\otimes\proj{\psi}$,
where $P_\cX$ is the projector onto $\cX$, gives an operator which is
proportional to the identity on $\cC$.  We do not elaborate on this
improvement any further, as it is a straightforward consequence of the
technique presented in~\cite{chrkoemire06}.
\end{remark} 
\begin{proof}
Let $\ket{\psi}\in\cB$ be arbitary and define
$\ket{\psi_\gel}=\gel\ket{\psi}$ for $\gel\in\cG$.  
Schur's Lemma in the form~\eqref{eq:squareintegrableprop} tells us
that the operator $d_{\cB}\int\proj{\psi_\gel}d\gel$ acts as the
identity on $\cB$ where integration is over the normalised Haar
measure on $\cG$. In particular, for $\ket{\Psi}\in\cC$, we have
\bea  \label{overlap}
\tr_\cB \proj{\Psi}&=&d_\cB \ \tr_{\cB}\left[\int (P_{\cA} \otimes \proj{\psi_\gel}) \proj{\Psi}d\gel\right]\nonumber\\
&=&\int \proj{\tilde{\chi}_{(\gel)}}dm(\gel)\label{eq:sumident}\ ,\\\nonumber
\eea
 where the normalised states $\ket{\tilde{\chi}_{(\gel)}}$ and the
probability measure $m$ on $\cG$ are defined by 
\begin{align}\label{eq:alphaphigdef}
\proj{\tilde{\chi}_{(\unitary )}} dm(\gel)=d_\cB\ \tr_{\cB}\bigl((P_{\cA}\otimes\proj{\psi_\gel})\proj{\Psi}\bigr)d\gel\ 
\end{align} 
and where $P_\cA$ is the identity on $\cA$. (To see that $\ket{\tilde{\chi}_{\unitary}}$ is indeed a pure state, observe that the rhs.~of~\eqref{eq:alphaphigdef} is of the form $d_{\cB} \tr_{\cB} \proj{\Phi_{(\gel)}}$, where $\ket{\Phi_{(\gel)}}=(P_\cA\otimes\proj{\psi_{\gel}})\ket{\Psi}$ is a product of pure states.)

Let $P$ be the projector onto $\cX$, and let $P_\gel=\gel
P\gel^\dagger$ be the projector onto $\gel\cX$. We claim that
$\ket{\tilde{\chi}_{(\gel)}}$ is on average close to the projected
state
\[
\ket{\chi_{(\gel)}}:=
\frac{P_\gel\ket{\tilde{\chi}_{(\gel)}}}{\bra{\tilde{\chi}_{(\gel)}}P_\gel\ket{\tilde{\chi}_{(\gel)}}}
\ .
\]
For this purpose, we use the gentle measurement lemma (see
e.g.,~\cite{winter99,ogawanagaoka02}), which implies with the triangle inequality that
\[
\bigl\|\proj{\tilde{\chi}_{(\gel)}}-\proj{\chi_{(\gel)}}\bigr\|\leq 2\sqrt{1-\bra{\chi_{(\gel)}}P_\gel\ket{\chi_{(\gel)}}}\ .
\]
We thus obtain 
\begin{align}
\bigl\|\tr_\cB\proj{\Psi}-\int\proj{\chi_{(\gel)}} dm(\gel)\bigr\|&\leq 2\int \sqrt{1-\bra{\tilde{\chi}_{(\gel)}}P_{\gel}\ket{\tilde{\chi}_{(\gel)}}}dm(\gel)\nonumber\\
&\leq 2\sqrt{1-\kappa}\ ,\ \label{eq:psidistbound}
\end{align}
using the convexity of the trace distance and the square root, 
where
\begin{align}\label{eq:deltapsidef}
\kappa=\int  \bra{\tilde{\chi}_{(\gel)}}P_{\gel}\ket{\tilde{\chi}_{(\gel)}}dm(\gel)\ .
\end{align}
Because of~\eqref{eq:alphaphigdef} and the cyclicity of the trace, we have
\begin{align*}
 \bra{\tilde{\chi}_{(\gel)}}P_{\gel}\ket{\tilde{\chi}_{(\gel)}}dm(\gel)&=d_\cB\ \tr\bigl((P_\gel\otimes\proj{\psi_\gel})\proj{\Psi}\bigr)d\gel\\
&=d_\cB\ \tr((P\otimes\proj{\psi})\proj{\Psi_{\gel^\dagger}})d\gel\ ,
\end{align*}
and therefore by the linearity and (again) the cyclicity of the trace
\begin{align*}
\kappa&=d_\cB\ \tr\bigl((P\otimes\proj{\psi})\int \proj{\Psi_{\gel^\dagger}} d\gel\bigr)\\
&=d_\cB\ \int \bra{\Psi}\gel P_\cC(P\otimes\proj{\psi})P_C\gel^\dagger \ket{\Psi} d\gel\ .
\end{align*}
Schur's Lemma (cf.~\eqref{eq:squareintegrableprop}) immediately
implies that $\kappa=\delta_{\ket{\psi}}(\cX)$. Since $\ket{\psi}$ was
arbitrary, we may take $\kappa=\delta(\cX)$ in
~\eqref{eq:psidistbound}, which concludes the proof.
\end{proof}

For later reference, we point out the following properties of the
quantity $\delta$.
\begin{lemma}\label{lem:deltaproperties}
Let $\cC \subset \cA \otimes \cB$, and let $\delta$ be as in
Definition~\ref{def:delta}. Then
\begin{enumerate}[(i)]
\item\label{eq:deltamonotony}
$\delta(\cX)\leq \delta(\cX')$ if $\cX \subset \cX'$.
\item\label{eq:deltamaximal} If $\cC$ is finite-dimensional,
$\delta(\cX)\leq 1$ with equality if $\cX=\cA$.
\end{enumerate}
\end{lemma}

\begin{proof}
Inequality~\eqref{eq:deltamonotony} follows directly from the
definition of $\delta(\cX)$.  Using the fact
that the projectors $P_{\cC}$ and $P_{\cA}$ are invariant, i.e., $\gel
P_{\cC}\gel^\dagger=P_{\cC}$ and $\gel P_{\cA}\gel^\dagger =P_{\cA}$,
we obtain using the cyclicity of the trace
\begin{align*}
 \tr[P_{\cC}(P_{\cA}\otimes\proj{\psi})]&= \tr[\gel^\dagger P_{\cC}\gel(P_{\cA}\otimes\proj{\psi})]\\
&= \tr[P_{\cC}(\gel P_{\cA}\gel^\dagger \otimes \proj{\psi_\gel})]\\
&= \tr[P_{\cC}(P_{\cA}\otimes\proj{\psi_\gel})]\ 
\end{align*}
for all $\ket{\psi}\in \cB$ and $\unitary\in \cG$. By linearity and Schur's lemma, we get
\begin{align}\label{eq:wdefidentity}
\tr(P_{\cC}(P_{\cA}\otimes \proj{\psi}))=\frac{\tr(P_{\cC}(P_{\cA}\otimes P_{\cB}))}{d_\cB}\ 
\end{align}
for any $\ket{\psi}$, which implies $\delta(\cA)=1$ by the definition
of $\delta(\cX)$. This proves~\eqref{eq:deltamaximal} when combined
with~\eqref{eq:deltamonotony}.
\end{proof}

\section{Exact expressions\label{sec:exactwstates}}

As a first application of the our theorem, we gives some examples
where $\delta(\cX)=1$. This implies that the approximation by
$\cX$-states in equation (\ref{eq:highweightapproximationerr}) is
exact. The subspaces $\cX$ in question are direct sums of weight
spaces for a particular range of weight values; we call these
$\cW^r$-states (see definition below). In this section, therefore, we
confine attention to semi-simple Lie groups, that have a weight space
structure.

\subsection{$\cW^r$-states for $\specialunitarygroup{d}$}

Recall first from Section~\ref{sec:gl} that $\heighthigh{\lambda}{\w}$
is defined as $\max(n_i)$, where $\lambda-\w=\sum_i n_i\alpha_i$ and
$\alpha_i$ are the simple roots.  Similarly, $\heightlow{\lambda}{\w}$
is defined as $\max(m_i)$, where $\w-\lambda_*=\sum_i m_i\alpha_i$,
and $\lambda_*$ is the lowest weight of $\cR_\lambda$. We will
partition the set of weights according to their height as follows.

\begin{definition} \label{Wdefinition}
Let $\cW_\lambda$ be the set of weights occuring in the irreducible representation $\cR_\lambda$ of $\specialunitarygroup{d}$. We define
$\weightshigh{\lambda}{r}$ for $r \ge 0$ to be the set of weights
$\weight{w}\in\weights{\lambda}$ satisfying
$\heighthigh{\lambda}{\weight{w}}\leq r$. Similarly,
$\weightslow{\lambda}{r}$ is the set of weights
$\weight{w}\in\weights{\lambda}$ satisfying
$\heightlow{\lambda}{\weight{w}}\leq r$. A
$\weightr{\lambda}{r}$-state is a state of the form $\gel\ket{\varphi}$, where $\gel\in\specialunitarygroup{d}$ and $\ket{\varphi}$ is supported on the weight space corresponding to $\weightslow{\lambda}{r}$.
\end{definition}
Thus $\weightshigh{\lambda}{r}$ consists of weights that lie within
distance $r$ of the highest weight, and $\weightslow{\lambda}{r}$
of those within distance~$r$ of the lowest weight.  Note that the same
set of states is generated by $\weightslow{\lambda}{r}$ and
$\weightshigh{\lambda}{r}$.  This is because $\specialunitarygroup{d}$
includes any permutation of the basis vectors. Thus
$\weightr{\lambda}{r}$ could equivalently have been defined in terms
of the high weights $\weightshigh{\lambda}{r}$.

We now consider states that can be expressed {\em exactly} as a convex
sum of $\weightr{}{r}$-states (we will often omit the representation
label when clear from the context).

\begin{corollary}[Exact expression by $\weightr{}{r}$-states] \label{cor:exact}
Let $\cR_\mu$ and $\cR_\nu$ be irreducible representations of
$\specialunitarygroup{d}$, and let $\cR_{\lambda}\subset\cR_\mu\otimes\cR_\nu$ be
a subrepresentation of their tensor product.  Then the partial trace
$\tr_{\cR_\nu} \proj{\Psi}$ of every state $\ket{\Psi} \in
\cR_\lambda$ is a convex sum of pure $\weightr{\mu}{r}$-states on
$\cR_\mu$, with
$r=\heightlow{\mu}{\weight{\lambda}-\weight{\nu}}$. That is, there is
a probability measure $m$ on $\specialunitarygroup{d}$ and states
$\ket{\chi_{(\unitary)}}=\gel\ket{\varphi_{(\gel)}}$ with
$\gel\in\specialunitarygroup{d}$ and $\ket{\varphi_{(\gel)}}$ supported on
the weight space
$\cR_\mu^{\cW^{\uparrow r}}$  such that
\begin{align}\label{eq:defbconv}
\tr_{\cR_\nu}\proj{\Psi}=\int\proj{\chi_{(\unitary )}} dm(\gel)\ .
\end{align}
\end{corollary}
\begin{proof}
We will derive this from Theorem~\ref{thm:highweight} by showing
that $\delta_{\ket{\weight{\nu}}}(\cW^{\uparrow r})=1$ for
$r=\heightlow{\mu}{\weight{\lambda}-\weight{\nu}}$, where $\ket{\weight{\nu}}$
is the highest weight vector in $\cR_\nu$. Let us define the set of
weights
\be \label{weight-range} \Lambda^{\uparrow}=
\{\weight{w} \in \weights{\mu} | \ \weight{w}+\weight{\nu} \in
\weights{\lambda}\}\ .  \ee 
Note that if $\weight{w}\in\Lambda^\uparrow$, then
$\weight{w}+\weight{\nu}\preceq \weight{\lambda}$ since
$\weight{\lambda}$ is the highest weight, and this is equivalent to
$\weight{w}\preceq\weight{\lambda}-\weight{\nu}$. We conclude
 that
$\Lambda^\uparrow\subset\weightslow{}{r}$, for
$r=\heightlow{\mu}{\weight{\lambda}-\weight{\nu}}$ since
$\heightlow{\mu}{\weight{w}}\leq \heightlow{\mu}{\weight{w}'}$
whenever $\weight{w}\preceq\weight{w}'$. By
Lemma~\ref{lem:deltaproperties}~\eqref{eq:deltamonotony}, it thus
suffices to show that
\begin{align}\label{eq:deltaeqone}
\delta_{\ket{\nu}}(\Lambda^\uparrow)=1\ .
\end{align}
Suppose first that $\ket{\Psi}\in\cR_\lambda$ is a weight vector of weight $\w_\lambda\in\weights{\cR_\lambda}$. Then $\ket{\Psi}$ has the form
\[
\ket{\Psi}=\sum_{\w_\mu,\w_\nu,j,k} \gamma_{jk}^{\w_\mu\w_\nu}\ket{\w_{\mu}}^{j}\ket{\w_{\nu}}^{k}
\]
where $\ket{\w_{\mu}}^{j}$ are weights vectors in $\cR_\mu$,
$\ket{\w_{\nu}}^{k}$ are weight vectors in $\cR_\nu$,
$\ket{\w_{\nu}}^{0}=\ket{\nu}$ is the highest weight vector in
$\cR_\nu$, and $\w_{\mu}+\w_\nu=\w_\lambda$ for all nonzero terms in the sum, by~\eqref{weight-expansion} and~\eqref{tensor-sum}. It is straightforward to check that
\begin{align}\label{eq:tracewa}
(P\otimes
\proj{\nu})\ket{\Psi}=(P_{\mu}\otimes\proj{\nu})\ket{\Psi}\ ,
\end{align}
where $P$ is the projector onto
the weightspace $\weightspace{\cR_\mu}{\Lambda^\uparrow}$, and where
$P_\mu$ is the identity on $\cR_\mu$.  But~\eqref{eq:tracewa} holds for all $\ket{\Psi}\in\cR_\lambda$, since $\ket{\Psi}$ can be expanded in terms of weight vectors.
This implies that 
\begin{align*}
\tr\bigl(\proj{\Psi}(P\otimes
\proj{\nu})\bigr)=\tr\bigl(\proj{\Psi}(P_{\mu}\otimes\proj{\nu})\bigr)\ .
\end{align*}
for all $\ket{\Psi}\in\cR_\lambda$. In particular, we have 
\begin{align*}
\delta_{\ket{\weight{\nu}}}(\Lambda^{\uparrow})&=\frac{d_\nu}{d_\lambda}\tr\bigl(P_\lambda(P_\mu\otimes\proj{\nu})\bigr)\\
 &=\frac{1}{d_\lambda}\tr\bigl(P_\lambda(P_\mu\otimes P_\nu)\bigr)\ ,
\end{align*}
where we used the invariance of $P_\lambda$, the cyclicity of the
trace and Schur's Lemma. Because
$\cR_\lambda\subset\cR_\mu\otimes\cR_\nu$, this
gives~\eqref{eq:deltaeqone}, as desired.
\end{proof}
The following lemma shows that it is natural to bound
$\Lambda^{\uparrow}$ by the set of weights $\weightslow{}{r}$:
\begin{lemma}\label{lowest}
The set $\Lambda^\uparrow$ (cf.~\eqref{weight-range}) includes the
lowest weight, $\mu_*$, of $\cR_\mu$. 
\end{lemma}
\begin{proof}(Compare \cite{knapp05}, Proposition 9.72.) Write any
weight vector $\ket{\w_\lambda}$ of $\cR_\lambda$ as $\ket{\w_\lambda}=\sum_{j}
\mu_{j}\ket{\w_{\mu,j}}\ket{\w_{\nu,j}}$ where the
sum is over pairs of weights $(\w_{\mu,j},\w_{\nu,j})$ of $\cR_\mu$
and $\cR_\nu$, respectively, satisfying
$\w_{\mu,j}+\w_{\nu,j}=\w_\lambda$ (abusing notation, in that we allow repetitions, if
necessary), and $\ket{\w_{\nu,j}}$ are weight vectors of $\cR_\nu$ and the
$\ket{\w_{\mu,j}}$ are mutually orthogonal weight vectors of $\cR_\mu$. Suppose
some weight $\w_{\nu,j}$ is maximal, in the sense that no
$\w_{\nu,j'}$ with $\w_{\nu,j'} \succ \w_{\nu,j}$ occurs in the
expansion. Apply the raising operator
$E_{rs}\in\mathfrak{sl}_d(\mathbb{C})$, i.e. the matrix that is zero
except for a $1$ in the $r$-th row and $s$-th column, with $r<s$. Then,
$E_{rs}\left(\ket{\w_{\mu,j}}\ket{\w_{\nu,j}}\right)=(E_{rs}\ket{\w_{\mu,j}})\ket{\w_{\nu,j}}+\ket{\w_{\mu,j}}(E_{rs}\ket{\w_{\nu,j}})$,
and the term $\ket{\w_{\mu,j}}(E_{rs}\ket{\w_{\nu,j}})$ cannot be
cancelled by some $\ket{\w_{\mu,{j'}}}(E_{rs}\ket{\w_{\nu,{j'}}})$ because
of the orthogonality of the $\ket{\w_{\mu,j}}$s, nor by some
$(E_{rs}\ket{\w_{\mu,j'}})\ket{\w_{\nu,j'}}$ because, if $E_{rs}\ket{\w_{\mu,j'}}$ has weight $\w_{\mu,j}$, then
$\w_{\mu,j'} \prec \w_{\mu,j}$, and hence $\w_{\nu,j'} \succ
\w_{\nu,j}$, in contradiction to the maximality of $\w_{\nu,j}$. Thus the term
$\ket{\w_{\mu,j}}(E_{rs}\ket{\w_{\nu,k}})$ can only vanish if killed by
$E_{rs}$, and it can only be killed by all raising operators if it is
the highest weight vector $\ket{\nu}$ (\cite{knapp05}, Theorem
5.5). So we must eventually, after a finite number of repeated
raisings, reach $\ket{\nu}$, showing that there is a weight vector of
$\cR_\lambda$ whose weight is $\w+\nu$ for some $\w \in
\weights{\mu}$. Starting with this vector and applying the same
procedure using lowering operators, and with the roles of $\mu$ and
$\nu$ interchanged, we must reach a weight vector with weight
$\mu_*+\nu$, and hence $\mu_* \in \Lambda^\uparrow$.
\end{proof}
Because of this Lemma, it makes sense to measure the distance $r$ from
the lowest weight $\mu_*$, as one implicitly does in defining
$r=\heightlow{\mu}{\weight{\lambda}-\weight{\nu}}$. Note that, if one
defines
\begin{align*} \label{weight-rangesecond}
\Lambda^\downarrow= \{\weight{w} \in \weights{\mu} | \
\weight{w}+\weight{\nu}_* \in \weights{\lambda}\}\ , 
\end{align*}
and sets $r=\heighthigh{\mu}{\weight{\lambda}_*-\weight{\nu}_*}$, one
obtains the same value of $r$ and the same approximating states as
those given by the theorem. This is because the map $w_i\leftrightarrow w_{d-i}$ on the components of a weight $\w$
interchanges the two sets of definitions.

We conclude with a simple example:
\begin{example}\label{example2}
Take $d=2$, $\mu=(k)$, $\nu=(n-k)$, where $k\leq n-k$ and  $\lambda=(n-\ell,\ell)\cong (n-2\ell)$.  Note that $k\geq \ell$ by the Littlewood-Richardson-rule. 
As the lowest weight $\mu_*$ is $(0,k)$, we have 
\begin{align*}
\lambda-\nu=(k-\ell,\ell)=\mu_*+(k-\ell)\alpha_1\ ,
\end{align*}
so $r=k-\ell$.

Rephrased in terms of angular momentum, we have $(k)=(2j_1)$,
$(n-k)=2j_2$ and $(n-2\ell)=(2j)$ and thus
$\heightlow{\weight{\mu}}{\weight{\lambda}-\weight{\nu}}=j_1-j_2+j$.
In particular, if $j\approx j_2-j_1$, then only a small number $r$ is
needed to obtain an exact expression in terms of $\cW^r$-states (cf.~Figure~\ref{fig:secondclebschgordanfigure}).
\end{example}

\subsection{$\cW^r$-states and exact expression for semi-simple Lie groups\label{sec:semi-simple}}

The results of the preceding section can be extended, almost without
change, to a simply-connected semi-simple Lie group $\cG$. Let
$\mathfrak{g}$ be the Lie algebra corresponding to $\cG$. A
representation of $\mathfrak{g}$ is a map from $\mathfrak{g}$ into
$\mathfrak{gl}(\cV)$ that preserves the Lie bracket. For a
simply-connected Lie group, the representations of $\cG$ and
$\mathfrak{g}$ are in one-to-one correspondence via the differential
map. This allows one to deal with the algebra $\mathfrak{g}$ rather
than the group $\cG$, which is often more convenient. 

Let $\mathfrak{h}$ be the Cartan subalgebra of $\mathfrak{g}$, and
suppose $\cR$ is a representation of $\mathfrak{g}$. Then a weight
vector $\ket{\weight{w}}\in\cR$ is a vector with the property that
$h\ket{\weight{w}} =\weight{w}(h) \ket{\weight{w}}$ for all $h\in
\mathfrak{h}$ where the {\em weight} $\weight{w}:\mathfrak{h}
\rightarrow \mathbb{C}$ is a linear functional. The {\em adjoint
representation}, $ad:\liealgebra\rightarrow
\mathfrak{gl}(\liealgebra)$, is defined by $ad(g)(h)=[g,h]$ for all
$h\in\liealgebra$. The weights of the adjoint representation are
called {\em roots}, and a subset $\{\alpha_1,\ldots,\alpha_d\}$ of
these is called a set of {\em simple roots} if every root $\beta$ can
be written uniquely as $\beta=\sum_{i=1} ^dn_i\alpha_i$, with integers
$(n_1,\ldots,n_d)$ which are either all nonnegative or all
nonpositive. A set of simple roots is a basis of the space
$\mathfrak{h}^*$ of linear functionals on $\mathfrak{h}$. The set of
weights is ordered by the rule that $\weight{w}\preceq \weight{w}'$ if
and only if $\weight{w}'-\weight{w}=\sum_{i=1} ^d n_i\alpha_i$ for
nonnegative $n_i\in\mathbb{R}$. With this rule, every irreducible
representation $\cR_\lambda$ is characterized by its highest weight
$\lambda$.

As with $\specialunitarygroup{d}$, every weight $\w$ of $\cR_\lambda$ can be
written as $\w=\lambda - \sum n_i \alpha_i$, and this enables us to
define $ht^{\downarrow}_\lambda(\w):=\max_i |n_i|$. Similarly, we can
define $ht^{\uparrow}_\lambda(\weight{w}):=\max_i |m_i|$, where
$\w=\lambda_*+\sum m_i\alpha_i$. We can now define $\cW^r$-states
exactly as in Definition~\ref{Wdefinition}, with $\cG$ replacing
$\specialunitarygroup{d}$, and we have:
\begin{corollary}[Exact expression by $\weightr{}{r}$-states for Lie groups] \label{cor:generalexact}
Let $\cR_\mu$ and $\cR_\nu$ be irreducible representations of a
 simply-connected semi-simple Lie group $\cG$, and let
$\cR_{\lambda}\subset\cR_\mu\otimes\cR_\nu$ be a subrepresentation of
their tensor product.  Then the partial trace $\tr_{\cR_\nu}
\proj{\Psi}$ of every state $\ket{\Psi} \in \cR_\lambda$ is a convex
sum of pure $\weightr{\mu}{r}$-states on $\cR_\mu$, with
$r=\heightlow{\mu}{\weight{\lambda}-\weight{\nu}}$. 
\end{corollary}
Lemma \ref{lowest} also carries over, if one replaces the raising
(lowering) operator $E_{rs}$ by a positive (negative) root of
$\mathfrak{g}$.

\section{Approximation using weight spaces}

\subsection{The case $\lambda=\mu+\nu$ for $\specialunitarygroup{d}$, and some numerical examples for~$\specialunitarygroup{2}$} 
Let us apply Theorem~\ref{thm:highweight} to rederive the following
known result (\cite[Theorem II.2]{chrkoemire06}), which can be used to
prove a standard de Finetti theorem (cf.~\cite{chrkoemire06}).

\begin{corollary}[The case $\lambda=\mu+\nu$]\label{cor:highestweight}
Let $\cR_{\lambda}\subset\cR_\mu\otimes\cR_\nu$ be irreducible
representations of $\specialunitarygroup{d}$, with $\lambda=\mu+\nu$,
let $\ket{\mu}$ be the highest weight vector in $\cR_\mu$ and let
$\ket{\Psi}\in\cR_\lambda$. Then there is a probability measure $m$ on
$\specialunitarygroup{d}$ such that
\[
\bigl\|\tr_{\cR_\nu}\proj{\Psi}-\int \gel \proj{\mu}\gel^\dagger dm(\gel)\bigr\|\leq 2\bigl(1-\frac{d_{\cR_\nu}}{d_{\cR_\lambda}}\bigr)\ .
\]
\end{corollary}
\begin{proof}
This is directly obtained from Theorem~\ref{thm:highweight} by
computing  $\delta_{\ket{\nu}}(\cW^{\downarrow 0})$, where $\ket{\nu}$
is the highest weight vector in $\cR_\nu$ because the weight space corresponding to $\cW^{\downarrow 0}$ is spanned by $\ket{\mu}$. Since $\ket{\mu}\otimes\ket{\nu}$ is the highest-weight vector in $\lambda$, we get  $\delta_{\ket{\nu}}(\cW^{\downarrow 0})=\textfrac{d_{\cR_\nu}}{d_{\cR_\lambda}}$ which concludes the proof.
\end{proof}

Our theorem also allows us to extend this to representations with
highest weights $\lambda\neq \mu+\nu$ occuring in the tensor product
$\cR_\mu\otimes\cR_\nu$, and include more weights in the
approximation, i.e., use $\cW^r$-states for $r>0$.

\begin{figure}
\centerline{\includegraphics[scale=0.49,angle=270]{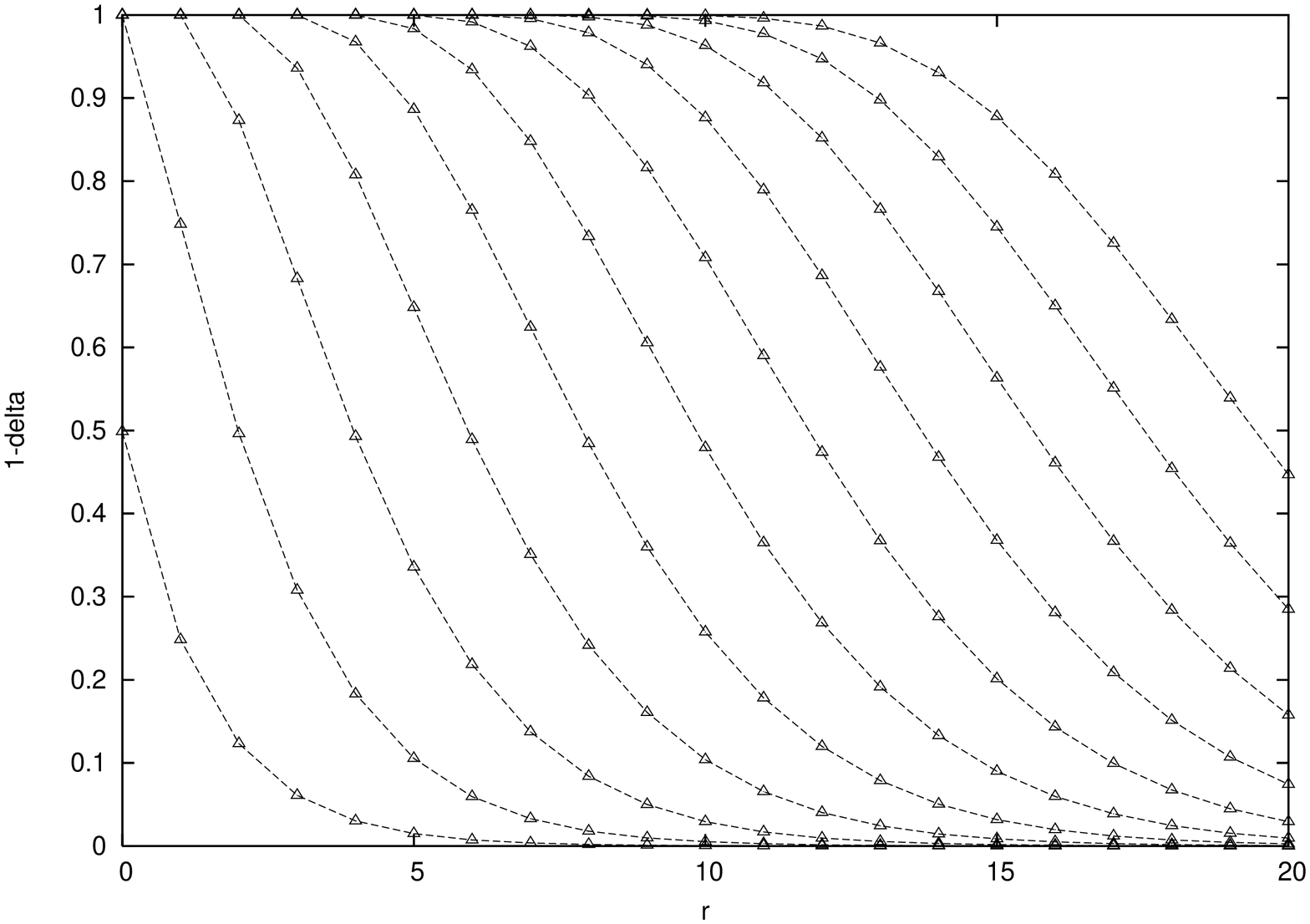}}
\caption{This figure shows the quantity
  $1-\delta_{\ket{j_2j_2}}(\cW^{\downarrow r})$
as a function of $r$. By Corollary~\ref{thm:highweight}, this quantity
is a measure of the error made when approximating
$\tr_{j_2}\proj{\Psi}$ for $\ket{\Psi}\in\cR_{(2j)}\subset \cR_{(2j_1)}\otimes \cR_{(2j_2)}$ by states with angular
momentum $m_1\geq j_1-r$ in some direction. The different curves
correspond to different values of $j$ in the regime where $j\approx
j_1+j_2$. The solid innermost curve is for $j=j_1+j_2$, and the dashed
outermost curve corresponds to total angular momentum
$j=j_1+j_2-10$. In this example, $j_1=j_2=100$.  The point with $r=0$ and
$1-\delta\approx 0.5$ of the innermost line corresponds to the result of~\cite{chrkoemire06} discussed at the end
of Section~\ref{sec:approximationwstates}, where the highest weight
vector $\ket{(j_1+j_2)(j_1+j_2)}=\ket{j_1j_1}\otimes\ket{j_2j_2}$ is a
product. This no longer holds for general $j$. We thus expect the
approximation to become worse for smaller $j$. The curves for smaller
total angular momentum $j$ show that we can nevertheless obtain a
small approximation error when we include high weights (i.e.,
$\cW^r$-states for $r>0$).
\label{fig:clebschgordanfigure}}
\end{figure}

\begin{figure}
\centerline{\includegraphics[scale=0.49,angle=270]{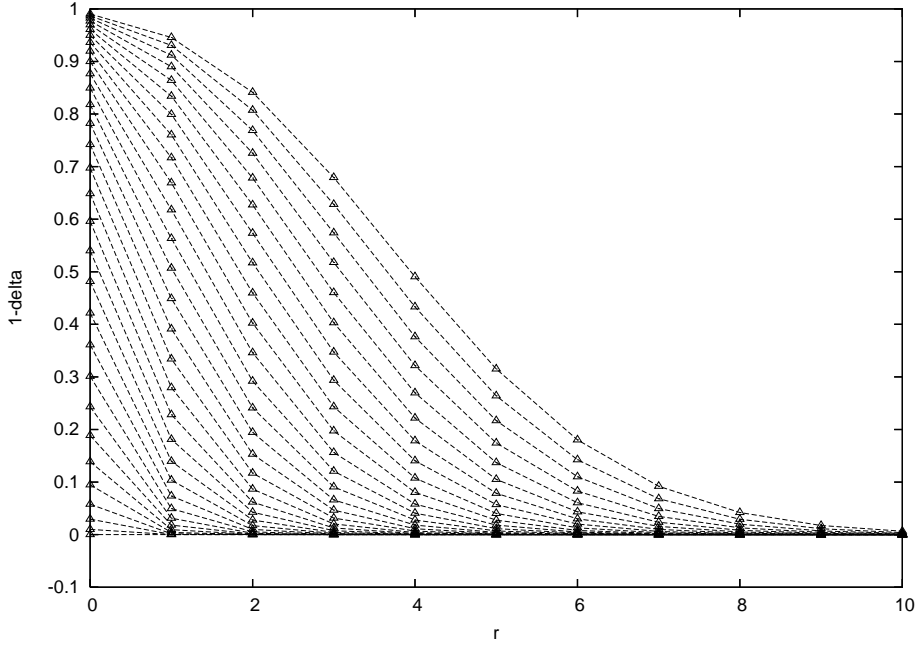}}
\caption{This figure shows $1-\delta_{\ket{j_2j_2}}(\cW^{\uparrow r})$
as a function of $r$ in the regime where $j\approx |j_1-j_2|$. More
precisely, we set $j_1=j_2=100$ and $j\in \{0,\ldots,30\}$ (the curve
corresponding to $j=0$ is the innermost one).  In accordance with
Corollary~\ref{cor:exact}, we can see from the figure that the error
made in the approximation by $\cW^r$-states
(cf.~Corollary~\ref{thm:highweight}) vanishes as soon as sufficiently
many high weights are used ($r\geq j$; see Example
\ref{example2}). Observe, however, that the error becomes exceedingly
small for $r$ significantly below the value $j$; e.g. for $j=30$
(right-most curve), where $1-\delta$ is close to zero for $r=10$.
\label{fig:secondclebschgordanfigure}}
\end{figure}

Next we apply Theorem~\ref{thm:highweight} to the case of
$\specialunitarygroup{2}$, where we can obtain bounds on the error
made in the approximation by using the Racah formula for the
Clebsch-Gordan coefficients to calculate $\delta(\cW)$.  That is, we
would like to approximate the partial trace $\tr_{\cR_{(2j_2)}}
\proj{\Psi}$ of a state $\ket{\Psi}\in \cR_{(2j)}\subset
\cR_{(2j_1)}\otimes \cR_{(2j_2)}$ with total angular momentum
$|j_1-j_2|\leq j\leq j_1+j_2$. Note that each of these representations
occurs with multiplicity $1$ and the case $\lambda=\mu+\nu$
corresponds to $j=j_1+j_2$. It is straightforward to check that the
weight space corresponding to the extremal weights $\cW^{\downarrow
r}$ is spanned by $\ket{j_1m_1}$ with $m_1\geq j_1-r$. These
correspond to states with high angular momentum in the $Z$-direction.

With Theorem~\ref{thm:highweight},  we conclude that there is a probability measure $m$ on $\specialunitarygroup{2}$ and states $\ket{\varphi_{(\gel)}}$ supported on eigenstates of the angular momentum in the $Z$-direction with eigenvalue at least $j_1-r$ such that 
\begin{align*}
\bigl\|\tr_{j_2}\proj{\Psi}-\int \gel \proj{\varphi_{(\gel)}}\gel^\dagger dm(\gel)\bigr\|\leq 2\bigl(1-\delta_{\ket{j_2m_2}}(\cW^{\downarrow r})\bigr)
\end{align*}
where $m_2\in\{-j,-j+1,\ldots,j\}$ is arbitrary and 
\begin{align}\label{eq:deltaclebschgordan}
\delta_{\ket{j_2m_2}}(\cW^{\downarrow r})=\frac{2j_2+1}{2j+1}\sum_{m_1=j_1-r} ^{j_1}
|\langle j(m_1+m_2)|j_1m_1j_2m_2\rangle|^2\ ,
\end{align}
as a straightforward calculation shows. An equivalent statement holds for $\cW^{\uparrow r}$. We give two numerical examples of this type of approximation in Figures~\ref{fig:clebschgordanfigure} and~\ref{fig:secondclebschgordanfigure}.

\subsection{Symmetric States: the exponential de Finetti Theorem\label{sec:symmetricstatesexp}}
We now focus on symmetric states
$\ket{\Psi}\in\Sym{\mathbb{C}^d}{n}$. Here, the {\em symmetric
subspace} $\Sym{\mathbb{C}^d}{n}\subset (\mathbb{C}^d)^{\otimes n}$ is
the subspace of vectors which are invariant under permutations of the
systems, i.e., under the action of the symmetric group $S_n$ on
$(\mathbb{C}^d)^{\otimes n}$ defined by
\begin{align}\label{eq:permutationaction}
\pi(\ket{\varphi_1}\otimes\cdots\otimes\ket{\varphi_n}):=
\ket{\varphi_{\pi^{-1}(1)}}\otimes\cdots\otimes\ket{\varphi_{\pi^{-1}(n)}}\ 
\end{align}
for all vectors
$\ket{\varphi_1},\ldots,\ket{\varphi_n}\in\mathbb{C}^d$ and $\pi\in S_n$. The group $\specialunitarygroup{d}$ acts on the same space simply by the $n$-fold
tensor product $\gel^{\otimes n}$ for every $\gel\in\specialunitarygroup{d}$,
i.e., its action is given by
\begin{align}\label{eq:unitaryaction}
\unitary(\ket{\varphi_1}\otimes\cdots\otimes\ket{\varphi_n}):=(\unitary\ket{\varphi_1})\otimes\cdots\otimes (\unitary\ket{\varphi_n})\ 
\end{align}
for all vectors
$\ket{\varphi_1},\ldots,\ket{\varphi_n}\in\mathbb{C}^d$ and
$\gel\in\specialunitarygroup{d}$. The
actions~\eqref{eq:permutationaction} and~\eqref{eq:unitaryaction}
commute, and $\Sym{\mathbb{C}^d}{n}$ is well known to be an
irreducible subspace with respect to the action of
$\specialunitarygroup{d}$. Its dimension is
$\dim\Sym{\mathbb{C}^d}{n}=\binom{n+d-1}{n}$.

The set of weights $\weights{\Sym{\mathbb{C}^d}{n}}$ occurring in the
symmetric representation is given by the set of all $d$-tuples
$\weight{w}=(w_1,\ldots,w_d)$ of nonnegative integers summing to
$n$. The weight space corresponding to a weight
$\weight{w}=(w_1,\ldots,w_d)\in\weights{\Sym{\mathbb{C}^d}{n}}$ is
one-dimensional and spanned by the vector
\begin{align}\label{eq:highestweightvecdef}
\ket{\weight{w}}=\frac{1}{\sqrt{|T^\weight{w}|}}\sum_{(v_1,\ldots,v_n)\in T^\weight{w}}\ket{v_1}\otimes\cdots\otimes\ket{v_n}\ .
\end{align}
In this expression, the {\em type class} $T^\weight{w}$ is defined as the set
of all $n$-tuples $\weight{w}$ where $i$ occurs $w_i$ times, for $1 \le i \le
d$. Thus the highest weight vector
$\ket{\weight{(n)}}=\ket{1}^{\otimes n}$ is of
product form.

For the symmetric representation, writing the weights as vectors of
integers, we have $\lambda=(n,0,\ldots,0)$, $\w=(w_1,\ldots, w_d)$,
and, since $\alpha_i=(0,0, \ldots, 1, -1, 0)$ with `1' in the $i$-th
position, we can write $n_i=n-\sum_{j=1}^{i}w_j$. So
$\heighthigh{\lambda}{\w}$ is just $n-w_1$, and our distance measure
just counts the number of times that a `1' in the highest weight
vector is replaced by some other number. We conclude that
$\cW^{\downarrow r}$ of $\Sym{\mathbb{C}^d}{n}$ is given by
\begin{align}\label{eq:weightsrsymmetric}
\cW^{\downarrow r}=\{\weight{w}\in\weights{\Sym{\mathbb{C}^d}{n}}\ |\ w_1\geq n-r\}\ .
\end{align}
Together with expression~\eqref{eq:highestweightvecdef}, this tells us
that the weight space corresponding to $\cW^{\downarrow r}$ is spanned
by states which are of the form $\ket{1}^{\otimes
n-r}\otimes\ket{\varphi}$ for (arbitrary) states
$\ket{\varphi}\in(\mathbb{C}^d)^{\otimes r}$ up to permutations of the
subsystems. Such almost-product states were called
``$\binom{n}{n-r}$-i.i.d. with prototype $\ket{1}$''
in~\cite{Ren07}. In this terminology, a $\cW^r$-state $\ket{\chi}$ is
a symmetric state in the subspace spanned by states which are
$\binom{n}{n-r}$-i.i.d. with prototype $\gel\ket{1}$, where
$\gel\in\specialunitarygroup{d}$ is determined by $\ket{\chi}$.

We can now state the main result of this section, which tells us about
the approximation of the partial trace of a symmetric state
$\ket{\Psi}\in\Sym{\mathbb{C}^d}{n}$ by $\cW^r$-states. A theorem of
this type was first proved in~\cite{Ren05}, and is further discussed
and given a somewhat simpler proof in~\cite{Ren07}.

\begin{corollary}[Exponential approximation by $\cW^r$-states]\label{cor:almostproduct}
Let $0\leq r\leq k\leq n$. Then the partial trace
$\tr_{n-k}\proj{\Psi}$ of every state
$\ket{\Psi}\in\Sym{\mathbb{C}^d}{n}$ can be approximated by a convex
combination of $\cW^r$-states in $\Sym{\mathbb{C}^d}{k}$ with error
\begin{align}\label{eq:almostproductdelta}
\varepsilon_{n,k,r,d}:=2\frac{\dim \Sym{\mathbb{C}^d}{n-k}}{\dim \Sym{\mathbb{C}^d}{n}}\sum_{i=r+1}^k\frac{\binom{k}{i}}{\binom{n}{i}}\binom{i+d-2}{i}\ .
\end{align}
That is, there is a probability
measure $m$ on $\specialunitarygroup{d}$ and states
$\ket{\chi_{(\unitary)}}=\gel \ket{\varphi_{(\gel)}}$, where $\gel\in\specialunitarygroup{d}$ and $\ket{\varphi_{(\gel)}}$ is supported on the weight space corresponding to $\cW^{\downarrow r}$,
such that 
\begin{align}
\bigl\|\tr_{n-k}\proj{\Psi}-\int\proj{\chi_{(\unitary )}} dm(\gel)\bigr\|\leq \varepsilon_{n,k,r,d}\ .
\end{align}
In particular, 
\begin{enumerate}[(i)]
\item\label{eq:combinatorialexactdefinetti}
if $r=0$, then 
$\varepsilon_{n,k,0,d}=2(1-\frac{\dim\Sym{\mathbb{C}^d}{n-k}}{\dim\Sym{\mathbb{C}^d}{n}})$.
\item\label{secndcombpropr}
if $d\leq \min\{k,n-k\}$, then 
$\varepsilon_{n,k,r,d}\leq 2\frac{e^{3d}}{(d-2)!} \Bigl(\frac{k}{n-r}\Bigr)^{r+1} \Bigl(\frac{k(n-k)}{n-r}\Bigr)^{d-2}$.
\end{enumerate}

\end{corollary}

\begin{proof}
We derive this result by applying Theorem~\ref{thm:highweight} to
$\Sym{\mathbb{C}^d}{n} \subset \Sym{\mathbb{C}^d}{k} \otimes
\Sym{\mathbb{C}^d}{n-k}$, and bounding $\delta_{\ket{\psi}}(\cW^{\downarrow r})$
for the highest weight vector $\ket{\psi}=\ket{1}^{\otimes n-k}$.

Consider a weight $\weight{w}\in\weights{\Sym{\mathbb{C}^d}{k}}$. It is easy to
verify for every permutation $\pi\in S_n$ and any two $v,v'\in T^\weight{w}$,
$\bra{v}\bra{1}^{\otimes n-k}\pi \ket{v'}\ket{1}^{\otimes n-k}$ is
either $1$ or $0$, and the number of permutations that map
$\ket{v}\ket{1}^{\otimes n-k}$ to $\ket{v'}\ket{1}^{\otimes n-k}$ is
equal to $(w_1+n-k)!\prod_{i=2} ^{d} w_i!$, for any $v$ and
$v'$ (Here, we write $\ket{v}$ for $\ket{v_1}\otimes\cdots\otimes\ket{v_k}$). 
In particular, since
$P_{\Sym{\mathbb{C}^d}{n}}=\frac{1}{n!}\sum_{\pi\in S_n} \pi$, we
conclude
\[
\tr\bigl(P_{\Sym{\mathbb{C}^d}{n}}(\ket{v'}\bra{v}\otimes \proj{1}^{\otimes n-k})\bigr)=\frac{1}{n!} (w_1+n-k)!\prod_{i=2} ^{d} w_i!\ .
\]
Using this identity and inserting the definition~\eqref{eq:highestweightvecdef}
of $\ket{\weight{w}}$, we obtain 
\begin{align}
\tr\bigl(P_{\Sym{\mathbb{C}^d}{n}}(\proj{\weight{w}}\otimes\proj{1}^{\otimes
  n-k})\bigr)&=\frac{|T^\weight{w}|}{n!}(w_1+n-k)!\prod_{i=2} ^{d} w_i!\nonumber\\ 
&=\frac{k!}{n!}\frac{(w_1+n-k)!}{w_1!}\ \label{eq:generalalpha}\ ,
\end{align}
where we inserted the cardinality
$|T^\weight{w}|=\frac{k!}{\prod_{i=1} ^{d} w_i!}$
of the type class. This implies  that for any set of weights $\cW\subset\weights{\Sym{\mathbb{C}^d}{k}}$, we have
\begin{align}\label{eq:deltapsiformula}
\delta_{\ket{\psi}}(\cW)=\frac{\dim \Sym{\mathbb{C}^d}{n-k}}{\dim
\Sym{\mathbb{C}^d}{n}}\frac{k!}{n!}\sum_{i=0} ^k \frac{(n-k+i)!}{i!}
f_i(\cW)\ ,
\end{align}
where $f_i(\cW)$ is the number of weights of $\cW$ with
$w_1=i$. Taking $\cW=\cW^r$, (\ref{eq:weightsrsymmetric}) tells us
that $f_i(\cW^r)=0$ unless $i \ge k-r$, and for any such $i$,
$f_i(\cW^r)$ counts all possible weights with $w_1=i$. This is just
the number of weights with symbols in
$\{2,\ldots,d\}$, or, equivalently, the dimension of $\Sym{d-1}{k-i}$,
which is $\binom{k+d-i-2}{k-i}$.  Thus we obtain
\begin{align}\label{eq:deltapsisecformula}
\delta_{\ket{\psi}}(\cW^r)=\frac{\dim \Sym{\mathbb{C}^d}{n-k}}{\dim
  \Sym{\mathbb{C}^d}{n}}\frac{k!}{n!}\sum_{i=k-r} ^k
\frac{(n-k+i)!}{i!} \binom{k+d-i-2}{k-i}\ .
\end{align}
However, we also know by
Lemma~\ref{lem:deltaproperties}~\eqref{eq:deltamaximal} that the sum
on the r.h.s., if extended to all $0\leq i\leq k$, is equal to $1$,
since this is the value
$\delta_{\ket{\psi}}(\weights{\Sym{\mathbb{C}^d}{k}})$. Combining this
with~\eqref{eq:deltapsisecformula} and substituting $i$ by $k-i$ then
shows that the expression~\eqref{eq:almostproductdelta} given in the
theorem is equal to $2(1-\delta_{\ket{\psi}}(\cW^r))$. The first claim
of the Theorem therefore follows from
Theorem~\ref{thm:highweight}. (Note that we can apply  Remark~\ref{rem:improvement} because the symmetric representation
occurs with multiplicity~$1$ in the tensor product
$\Sym{\mathbb{C}^d}{k}\otimes\Sym{\mathbb{C}^d}{n-k}$.)

We have already shown the bound~\eqref{eq:combinatorialexactdefinetti} for $r=0$ in Corollary~\ref{cor:highestweight}. The bound~\eqref{secndcombpropr} follows
from~\eqref{eq:almostproductdelta} after some further algebra that is
deferred to Appendix~\ref{app:combinatorial}.
\end{proof}

When $r=0$, only the highest weight is used, resulting in an
approximation by a convex combination of product states; this is the
standard de Finetti theorem. The bound we obtain in this case is given
as statement~\eqref{eq:combinatorialexactdefinetti}. It implies the
$2\frac{dk}{n}$ bound stated as Corollary~II.3 in~\cite{chrkoemire06}
and shown there to be optimal in the number of systems. The special
case where $k=2$ was recently treated
in~\cite{fannesvandenplas06}. Note that the version of the theorem in
\cite{Ren05,Ren07} does not yield a useful bound for the $r=0$ case;
this is because some algebraic steps lose precision.

When $r>0$, the approximation is by almost product states. The
bound~\eqref{secndcombpropr} explains why we call this an
exponential approximation. Consider for example the case where $d$
is fixed, with $k=\alpha n$ and $r=\beta k$ for some constants
$\alpha,\beta\in[0,1]$. In the limit as $n\rightarrow \infty$, the
bound has an asymptotic behaviour of the form $O(n^{d-2}
(e^{-\alpha\beta\log(\frac{1-\alpha\beta}{\alpha})})^n)$. This tends
to zero exponentially fast with $n$ for suitable parameters $\alpha$
and $\beta$.  Note that the bound $3(n-k)^d e^{-r(\frac{n-k}{n})}$
given in~\cite{Ren05} translates into
$O(n^d(e^{-\alpha\beta(1-\alpha)})^n)$, and thus the convergence of
our bound~\eqref{secndcombpropr} is better when $\beta\leq
\textfrac{1}{\alpha}-e^{1-\alpha}$, which is the case whenever
$\alpha\lesssim 0.34$.

By the arguments in~\cite{Ren05,chrkoemire06} we can extend this
theorem from pure states $\ket{\Psi}\in\Sym{\mathbb{C}^d}{n}$ to
general (mixed) states on $(\mathbb{C}^d)^{\otimes n}$ that are
symmetric, satisfying $\pi\rho^n\pi^\dagger=\rho^n$ for all
$\pi\in S_n$. We first purify $\rho^n$ in a symmetric way; this gives
a symmetric pure state $\ket{\Psi}\in
(\mathbb{C}^d\otimes\mathbb{C}^d)^{\otimes n}$ to which the theorem
can be applied. The approximation error is then given
by~\eqref{eq:almostproductdelta} with $d$ replaced by $d^2$, and the
approximating states are partial traces of pure almost product
states.

\section{An exponential theorem for the Heisenberg group\label{sec:heisenbergweyl}} 
We now turn to the Heisenberg group $\heis$. There is no weight space
structure for the representations of $\heis$, as the Heisenberg
algebra is nilpotent rather than semi-simple. However, the subspaces
spanned by a particular range of number states play the role
previously taken by weight spaces.

Consider the tensor product representation $\cH_\mu\otimes\cH_\nu$,
where $\cH_\mu$ and $\cH_\nu$ are irreducible representations with
parameters $\mu$, $\nu$ as in~\eqref{eq:lambdairrep} of
Section~\ref{sec:heisenbergweylintro}. For simplicity, we will
henceforth assume that both $\mu$ and $\nu$ are positive, but we point
out that our results can be extended to other cases.

 Our first aim is to
identify irreducible subspaces in the tensor product
$\cH_\mu\otimes\cH_\nu$. To do so, we will use the
realisation of $\cH_\mu$ and $\cH_\nu$ based on a pair of creation-
and annihilation operators $(a,a^\dagger)$
described
by~\eqref{eq:hlambdaaction}.  We will write $a_1=a\otimes
\id_{\cH_\nu}$ and $a_2=\id_{\cH_\mu}\otimes a$ and similarly for
$a_1^\dagger$ and $a_2^\dagger$. We then have commutation relations such
as $[a_1,a_2]=0$. By definition of the tensor product
representation, the element $(\alpha;t)\in\heis$ acts on
$\cH_\mu\otimes\cH_\nu$ as the operator
\begin{align}\label{eq:tensoractionweylheisenberg}
e^{i\mu t}D_{a_1}(\sqrt{\mu}\alpha)\otimes e^{i\nu t}
D_{a_2}(\sqrt{\nu}\alpha)=e^{i(\mu+\nu)t}
D_{a_{\mu\otimes\nu}}(\sqrt{\mu+\nu}\alpha)\ ,
\end{align}
where we used the commutation relations and identity~\eqref{eq:displacementoperatordef} and introduced the
operators
\begin{align}
a_{\mu\otimes\nu} =\frac{1}{\sqrt{\mu+\nu}}
(\sqrt{\mu}a_1+\sqrt{\nu}a_2)\qquad \textrm{and} \qquad 
a_{\mu\otimes\nu}^\dagger =\frac{1}{\sqrt{\mu+\nu}}
(\sqrt{\mu}a_1^\dagger+\sqrt{\nu}a_2^\dagger)\ .\label{eq:tensorprodcreatanni}
\end{align}
 It is straightforward to check that $a_{\mu\otimes\nu}^\dagger$ and
$a_{\mu\otimes\nu}$ satisfy canonical commutation
relations. Combining~\eqref{eq:tensoractionweylheisenberg}
with~\eqref{eq:hlambdaaction} thus demonstrates part~\eqref{it:subspc}
of the following:
\begin{lemma}\label{lem:subspacelemma}
Let $\mu,\nu>0$. Consider the  tensor product
representation $\cH_\mu\otimes\cH_\nu$ of the Heisenberg group $\heis$, where  $\cH_\mu$ and
$\cH_\nu$ are the irreducible representations as described
by~\eqref{eq:hlambdaaction}, and let $a_{\mu\otimes\nu}^\dagger$ and
$a_{\mu\otimes\nu}$ be defined by~\eqref{eq:tensorprodcreatanni}. Then
the following holds.
\begin{enumerate}[(i)]
\item\label{it:subspc}
Let $\ket{0}_{\mu\otimes\nu}\in\cH_\mu\otimes\cH_\nu$ be a normalised
vector such that $a_{\mu\otimes\nu}\ket{0}_{\mu\otimes\nu}=0$.  Then
the action of $\heis$ on $\ket{0}_{\mu\otimes\nu}$ generates an
irreducible subspace isomorphic to $\cH_{\mu+\nu}$ with
orthonormal basis
$\{\frac{(a_{\mu\otimes\nu}^\dagger)^{n}}{\sqrt{n!}}\ket{0}_{\mu\otimes\nu}\}_{n\in\mathbb{N}_0}$.
\item\label{it:killedvector}
Let $\ket{n}:=\frac{(a^\dagger)^n}{\sqrt{n!}}\ket{0}$ and define 
\[
\ket{\psi^\Delta}=\sum_{\ell=0} ^\Delta (-1)^\ell \sqrt{\alpha_\ell}
\ket{\Delta-\ell}\otimes\ket{\ell}\qquad\textrm{where }\qquad \alpha_\ell=\frac{\binom{\Delta}{\ell}\mu^\ell
  \nu^{\Delta-\ell}}{(\mu+\nu)^\Delta}
\]
for $\Delta\in\mathbb{N}_0$. Then $\ket{\psi^\Delta}$ is a normalised
vector in $\cH_\mu\otimes\cH_\nu$ with
$a_{\mu\otimes\nu}\ket{\psi^\Delta}=0$.
\item\label{it:formaldegreest}
Let $\cH^\Delta_{\mu+\nu}\subset\cH_\mu\otimes\cH_\nu$ be the irreducible
representation generated by $\ket{\psi^\Delta}$. Then
$\cH^\Delta_{\mu+\nu}$ is isomorphic to $\cH_{\mu+\nu}$ and has formal
degree $d_{\cH^\Delta_{\mu+\nu}}=\mu+\nu$ as a
representation of the quotient group $\heis/Z$.
\item\label{it:completedecomposition}
The tensor product representation decomposes into a direct sum of the
representations $\cH^\Delta_{\mu+\nu}$; that is,
\begin{align*}
\cH_\mu\otimes\cH_\nu\cong\bigoplus_{\Delta\in\mathbb{N}_0} \cH^\Delta_{\mu+\nu}\ .
\end{align*}
\end{enumerate}
\end{lemma} 
\begin{proof}
Statement~(\ref{it:killedvector}) follows by straightforward
computation using~\eqref{eq:tensorprodcreatanni}.
Statement~\eqref{it:formaldegreest} is a consequence of the fact that
$d_{\cH_{\mu+\nu}}=\mu+\nu$ (cf.~end of the
Section~\ref{sec:heisenbergweylintro}) and the fact that the
representation $\cH_{\mu+\nu}$ is identical to $\cH_{\mu+\nu}^\Delta$
when the ``vacuum state'' $\ket{0}$ is identified with
$\ket{\psi^\Delta}$ and the operators $a$, $a^\dagger$ are identified
with~\eqref{eq:tensorprodcreatanni}.  To
prove~\eqref{it:completedecomposition}, we first prove that the
subspaces $\cH_{\mu+\nu}^\Delta$ and $\cH_{\mu+\nu}^{\Delta'}$ are
orthogonal for $\Delta\neq \Delta'$. By definition, this is equivalent
to showing $\spr{\varphi^\Delta_n}{\varphi^{\Delta'}_{n'}}=0$ for all
$n$ and $n'$, where
$\ket{\varphi^\Delta_n}:=(a_{\mu\otimes\nu}^\dagger)^n\ket{\psi^\Delta}$.
Because $\ket{\varphi^\Delta_n}$ is supported on the span of the
states $\{\ket{n_1}\otimes\ket{n_2}\ |\ n_1+n_2=n+\Delta\}$, this
certainly holds when $\Delta+n\neq \Delta'+n'$. If on the other hand
$\Delta+n= \Delta'+n'$, then the overlap between
$\ket{\varphi^\Delta_n}$ and $\ket{\varphi^{\Delta'}_{n'}}$ must
vanish since these two vectors are eigenvectors of the hermitian
operator $a_{\mu\otimes\nu}^\dagger a_{\mu\otimes \nu}$ with distinct
eigenvalues $n\neq n'$. Finally, to prove that these subspaces provide
a complete decomposition of $\cH_\mu\otimes\cH_\nu$, it suffices to
observe that for each $m\in\mathbb{N}_0$, the subspace spanned by
$\{\ket{k}\otimes\ket{\ell}\ |\ k+\ell=m\}$ is identical to the span
of $\{\ket{\varphi^\Delta_n}\ |\ \Delta+n=m\}$.
\end{proof}

\begin{figure}
\centerline{\includegraphics[scale=0.49,angle=270]{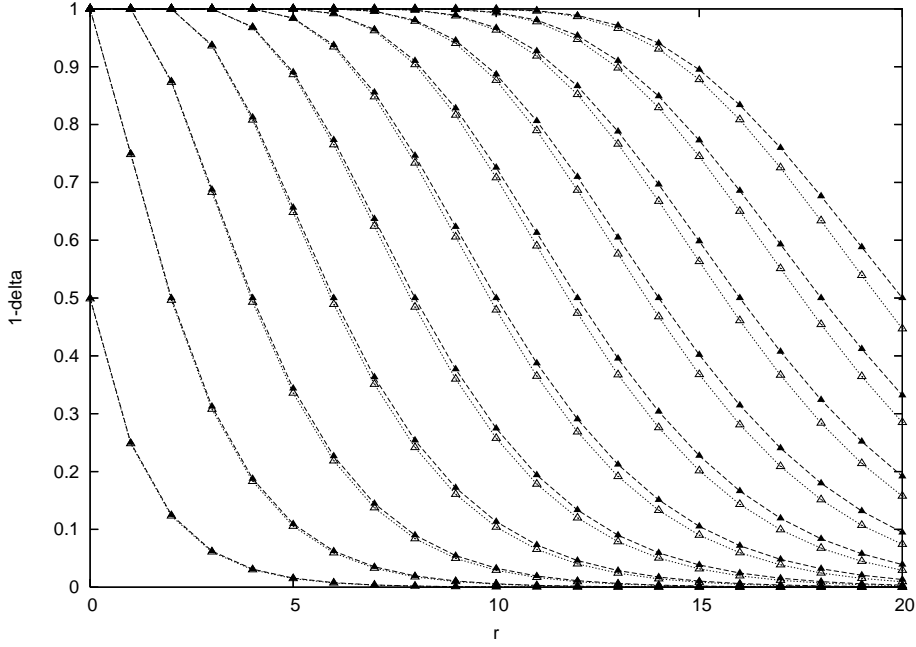}}
\caption{We illustrate the result of Corollary~\ref{cor:heisenbergweyl}
  for the case of the irreducible representations $\cH^\Delta_n\subset
  \cH_k\otimes\cH_{n-k}$ for $k=50$ and $n=100$. 
The figure shows the quantity $1-\delta_{\ket{0}}(\cN^r)$ as
  a function of $r$, where the different curves correspond to
  different values of $\Delta$ (dashed lines with filled
  triangles). The innermost curve corresponds to
  $\Delta=0$,  whereas the outermost curve corresponds to
  $\Delta=10$.  For comparison, the curves from
  Figure~\ref{fig:clebschgordanfigure} are also shown (dotted lines
  with empty triangles). Note how close they are, even though they
 represent very different objects: For
  $\specialunitarygroup{2}$, the curves correspond to inequivalent
  irreducible representations, whereas for the Heisenberg group, they
  are associated with different copies of the same representation in
  the tensor product.\label{fig:heisenbergweylfigure}}
\end{figure}

We will now study the partial trace of a state
$\ket{\Psi}\in\cH^\Delta_{\mu+\nu}$ and show that it can be
approximated by rotated number states. For an irreducible
representation $\cH_\lambda$, we set $\cN_{\cH_\lambda}=\mathbb{N}_0$,
and, for a set of numbers $\cN\subset\cN_{\cH_\lambda}$, we will
define the corresponding number subspace $\cH_\lambda^\cN$ as the span
of the states $\ket{n}$, $n\in\cN$ obtained by raising the vacuum
state $n$ times. By analogy with the set of extremal weights
$\weightslow{\lambda}{r}$ we define, for $r\in\mathbb{N}_0$,
\[
\cN^r:=\{n\ |\ n\leq r\}\subset \cH_\mu\ .
\]
 The error term in the following corollary depends on the way the
representation $\cH_{\mu+\nu}$ is embedded into
$\cH_\mu\otimes\cH_\nu$; different values of $\Delta$ lead to a
different approximation error.
\begin{corollary}[Number space approximation for the Heisenberg
  group]\label{cor:heisenbergweyl}
Let $\mu,\nu>0$ and let $\Delta\in\mathbb{N}_0$.
Let $\cH^\Delta_{\mu+\nu}\subset\cH_\mu\otimes\cH_\nu$ be irreducible representations of the Heisenberg group as described
by Lemma~\ref{lem:subspacelemma}. Then 
\begin{align}\label{eq:deltaheisenbergweyl}
\delta_{\ket{0}}(\cN^r)&=\bigl(\frac{\nu}{\mu+\nu}\bigr)^{\Delta+1}\sum_{n=0}
^{r-\Delta}\binom{n+\Delta}{\Delta}
\bigl(\frac{\mu}{\mu+\nu}\bigr)^n\ .
\end{align}
In particular,  for every $\ket{\Psi}\in\cH^\Delta_{\mu+\nu}$, the
partial trace $\tr_{\cH_\nu}\proj{\Psi}$ can be approximated by a
convex combination of $\cN^r$-states. That is, there are states $\ket{\chi_{(\alpha)}}$ and a probability measure $m$ on $\mathbb{C}$ such that 
\begin{align*}
\bigl\|\tr_{\cU_\nu}\proj{\Psi}-\int \proj{\chi_{(\alpha)}} dm(\alpha)\bigr\|\leq \varepsilon_{\mu,\nu,\Delta,r}\ ,
\end{align*}
where  $\ket{\chi_{(\alpha)}}$ is supported on $D_{a_1}(\alpha)\
\myspan\{\ket{n}\ |\ n\leq r\}$ and where
\begin{align*}
\varepsilon_{\mu,\nu,\Delta,r}=\begin{cases}
2\bigl(1-\delta_{\ket{0}}(\cN^r)\bigr)\qquad &\textrm{ if
  }\Delta=0\textrm{ and }r=0\\
2\sqrt{1-\delta_{\ket{0}}(\cN^r)}\qquad &\textrm{otherwise}\ .
\end{cases}
\end{align*}
In particular, if $\Delta=0$, then 
\begin{align*}
\varepsilon_{\mu,\nu,0,r}=\begin{cases}
2\bigl(\frac{\mu}{\mu+\nu}\bigr)\qquad &\textrm{ if
  } r=0\\
2\bigl(\frac{\mu}{\mu+\nu}\bigr)^{\textfrac{(r+1)}{2}}\qquad &\textrm{ otherwise }\ .
\end{cases}
\end{align*}

\end{corollary}
We use this corollary to produce Figure~\ref{fig:heisenbergweylfigure},
which shows both the exponential decay of the approximation error when
more approximating states are used (i.e., for varying $r$) and how
different embeddings of the same representation $\cH_{\mu+\nu}$ into
the tensor product $\cH_\mu\otimes\cH_\nu$ (i.e., different values of
$\Delta$) give rise to varying  approximation errors.

\begin{proof}
We will prove identity~\eqref{eq:deltaheisenbergweyl}. The result
then is a direct consequence of  Theorem~\ref{thm:highweight} and Remark~\ref{rem:improvement}.

The vectors
$\ket{\psi^\Delta_n}=\frac{(a_{\mu\otimes\nu}^\dagger)^n}{\sqrt{n!}}\ket{\psi^\Delta}$
are a basis of the subspace $\cH_{\mu+\nu}^\Delta$ generated by $\ket{\psi^\Delta}$ according
to Lemma~\ref{lem:subspacelemma}. Because of the commutativity of
$a_1^\dagger$ and $a_2^\dagger$, we have
\begin{align*}
(a_{\mu\otimes\nu}^\dagger)^n=\bigl(\frac{1}{\sqrt{\mu+\nu}}\bigr)^{n}\sum_{s=0}
^n
\binom{n}{s}\sqrt{\mu}^s\sqrt{\nu}^{n-s}(a_1^\dagger)^s(a_2^\dagger)^{n-s}\ .
\end{align*}
Note also that
$(a^\dagger)^s\ket{\ell}=\frac{\sqrt{(s+\ell)!}}{\sqrt{\ell!}}\ket{s+\ell}$.
In particular, (by setting $s=n$ and $\ell=0$) we obtain
\[
\ket{\psi^\Delta_n}=\sqrt{\alpha_{\Delta,n}}\ket{\Delta+n}\otimes \ket{0}+\ldots\ ,
\]
where \begin{align}\label{eq:alphadeltandef}
\alpha_{\Delta,n}=\bigl(\frac{\nu}{\mu+\nu}\bigr)^\Delta \binom{n+\Delta}{\Delta}\bigl(\frac{\mu}{\mu+\nu}\bigr)^n\ 
\end{align}
and the remaining terms  are supported on $\cH_\mu\otimes
\myspan\{\ket{n}\ |\ n\geq 1\}$. We can then compute
\begin{align}\label{eq:sumid}
\sum_{n=0} ^\infty\sum_{n'=0} ^r \tr\bigl(\proj{\psi^\Delta_n}
(\proj{n'}\otimes\proj{0})\bigr)&=\sum_{n=0} ^{r-\Delta}
\alpha_{\Delta,n}\ .
\end{align}
Identity~\eqref{eq:deltaheisenbergweyl} now follows
from~\eqref{eq:alphadeltandef} and~\eqref{eq:sumid} using 
the definition of $\delta_{\ket{0}}(\cN^r)$ and the formal degrees
$d_{\cH_\nu}=\nu$ (cf. discussion at the end of
Section~\ref{sec:heisenbergweylintro}) and
$d^\Delta_{\cH_{\mu+\nu}}=\mu+\nu$ (Lemma~\ref{lem:subspacelemma}). 
\end{proof}
Corollary~\ref{cor:heisenbergweyl} turns into the result
of~\cite{CruzOsborneSchack06} when specialised to integer $\mu$ and
$\nu$, and $r=\Delta=0$. We will now explain why this is the case, and
how the subspace spanned by $k$-fold coherent states considered
in~\cite{CruzOsborneSchack06} is related to representations of the
Heisenberg group.

We begin by taking the $k$-fold tensor product of the irreducible
representation $\cH_1$. As before, let $a_i^\dagger$ and $a_i$ be the
creation and annihilation operators acting on the $i$-th factor in
this tensor product. As with~\eqref{eq:tensorprodcreatanni}, it
is then easy to see that the operators
\begin{align}
a=\frac{1}{\sqrt{k}}\sum_{i=1}^k a_i\qquad \textrm{and} \qquad 
a^\dagger=\frac{1}{\sqrt{k}}\sum_{i=1}^k a_i^\dagger\ .
\end{align}
satisfy canonical commutation relations, and the vector
$\ket{0}^{\otimes k}$ is annihilated by $a$. If we let $(\alpha;t)$
act as $e^{it}D_a(\alpha)$ on this vector, this generates an
irreducible representation isomorphic to $\cH_1$ with orthonormal
basis vectors
\begin{align}\label{eq:nkdfe}
\ket{n}_k=\frac{(a^\dagger)^n}{\sqrt{n!}}\ket{0}^{\otimes
k}=k^{-\textfrac{n}{2}}\sum_{(f_1,\ldots,f_k): \sum_i
f_i=n}\sqrt{\frac{n!}{f_1!\cdots f_k!}} \ket{f_1}\otimes\cdots\otimes
\ket{f_k}\ ,
\end{align}
where the sum is over all $k$-tuples $(f_1,\ldots,f_k)$ of nonnegative
integers. This is the space spanned by vectors of the form
$\{\ket{\varphi_\alpha}^{\otimes k}\}_{\alpha\in\mathbb{C}}$, where
$\ket{\varphi_\alpha}=e^{-\textfrac{|\alpha|^2}{2}}\sum_{n=0}^\infty\frac{\alpha^n}{\sqrt{n!}}\ket{n}$
is a {\em coherent} state. By the recipe described in
Section~\ref{sec:heisenbergweylintro}, we can use this to define an
irreducible representation isomorphic to $\cH_k$. This is done by
letting the element $(\alpha;t)$ act on the same space as
$e^{ikt}D_a(\sqrt{k}\alpha)$. If we call the resulting representation
$\cH^{(k)}$, then clearly
$\cH^{(n)}\subset\cH^{(k)}\otimes\cH^{(n-k)}$ as representations, and
this corresponds to the irreducible representation $\cH_n^0\subset
\cH_k\otimes\cH_{n-k}$ identified in
Lemma~\ref{lem:subspacelemma}. With
Corollary~\ref{cor:heisenbergweyl}, we thus obtain the following
generalised form of the statement of~\cite{CruzOsborneSchack06}.

\begin{corollary}\label{cor:coherentstate}
Let $a$ and $a^\dagger$ be operators satisfying $[a,a^\dagger]=\id$
and let $\ket{0}$ be such that $a\ket{0}=0$. Define the number states
$\ket{n}=\frac{(a^\dagger)^n}{\sqrt{n!}}\ket{0}$ for
$n\in\mathbb{N}_0$ and the displacement operators
$D(\alpha)=\exp(\alpha a^\dagger-\bar{\alpha}a)$ for
$\alpha\in\mathbb{C}$. Finally, let
$\ket{\Psi}\in\Sym{\cH}{n}\subset\cH^{\otimes n}$ be a pure state in
the span of the set of states of the form $(D(\alpha)\ket{0})^{\otimes
n}$. Then there are states $\ket{\chi_{(\alpha)}}$ and a probability measure $m$ on $\mathbb{C}$ such that 
\begin{align*}
\bigl\|\tr_{n-k}\proj{\Psi}-\int
\proj{\chi_{(\alpha)}}dm(\alpha)\bigr\|\leq\begin{cases}
2\frac{k}{n}\qquad &\textrm{ if }r=0\\
2\left(\frac{k}{n}\right)^{\textfrac{(r+1)}{2}} \qquad &\textrm{otherwise,}
\end{cases}
\end{align*}
where $\ket{\chi_{(\alpha)}}$ is supported on $D(\alpha)^{\otimes k}
\myspan\{\ket{n}_k\ |\ n\leq r\}$, with $\ket{n}_k$ defined by
~\eqref{eq:nkdfe}. In particular, for $r=0$ we have
$\ket{\chi_{(\alpha)}}=(D(\alpha)\ket{0})^{\otimes k}$.
\end{corollary}

\section{Conclusions}
It is striking that representations lend themselves so well to
studying de Finetti theorems. Schur's lemma is the essential
representation-theoretic tool in proving our main theorem. In
addition, for $\specialunitarygroup{d}$, and more generally for semi-simple groups, the
weight space structure provides a natural family of approximating
states that we call $\cW^r$-states; these amount, in the case of a
symmetric representation, to almost-product states. For the Heisenberg
group, subspaces $\cN^r$ spanned by sets of number states play the same role.

Our theorem assumes that we have irreducible representations $\cA$,
$\cB$ and $\cC$ satisfying $\cC \subset \cA \otimes \cB$, and then
tells us how well the trace $\tr_{\cB}\proj{\Psi}$ of a state
$\ket{\Psi}$ in $\cC$ can be approximated by $\cW^r$-states in $\cA$
(or $\cN^r$-states for the Heisenberg group). The quality of this
approximation is determined by the way $\cC$ is embedded in the
product $\cA \otimes \cB$. This is captured by the number $\delta$,
given by Definition \ref{def:delta}. We give several examples of
explicit calculations of this number:

\begin{enumerate}[(i)]
\item
 For the embedding of symmetric subspaces $\cR_{(n)} \subset \cR_{(k)}
\otimes \cR_{(n-k)}$, $\delta(\cW^r)$ is given by
equation~\eqref{eq:deltapsisecformula}. This corresponds to the
exponential theorem for symmetric states proved in~\cite{Ren05,Ren07},
and we obtain bounds from~\eqref{eq:deltapsisecformula} that
reproduce, and in fact slightly improve upon, the results in those
papers.

\item In the case of the Heisenberg group, there is an irreducible
  representation $\cH_\mu$ for every non-zero parameter
  $\mu\in\mathbb{R}$, but only representations isomorphic to
  $\cH_{\mu+\nu}$ occur in the tensor product
  $\cH_\mu\otimes\cH_\nu$. The corresponding $\delta(\cN^r)$ is given
  by equation~\eqref{eq:deltaheisenbergweyl} and depends on how this
  representation is embedded (see
  Corollary~\ref{cor:heisenbergweyl}). This allows us to prove an
  exponential theorem (Corollary~\ref{cor:coherentstate}) that
  generalises the result in~\cite{CruzOsborneSchack06}.

\item The theorem for representations of the unitary group proved
in~\cite{chrkoemire06} corresponds to $\cR_{\mu+\nu} \subset \cR_\mu
\otimes\cR_\nu$, where each $\cR$ is a representation of
$\specialunitarygroup{d}$, and in this case $\delta(\cW^r)=\dim \cR_\nu/\dim
\cR_{\mu+\nu}$ for $r=0$ (see Lemma~\ref{lem:deltaproperties}), giving
the bound for Theorem~2.1 in~\cite{chrkoemire06}.

\item
For $\specialunitarygroup{2}$, the mapping of weight vectors in $\cC$
into those of $\cA \otimes \cB$ is given by the Clebsch-Gordan
coefficients, and in that case the explicit expression for
$\delta(\cW^r)$ in equation~\eqref{eq:deltaclebschgordan} enables us
to give give bounds for the de Finetti approximation by $\cW^r$-states
for a range of values of angular momenta and~$r$ (see
Figure~\ref{fig:clebschgordanfigure}).

\item Given representations of $\specialunitarygroup{d}$ satisfying
$\cR_\lambda \subset \cR_\mu \otimes \cR_\nu$, we find
$\delta(\cW^r)=1$ when
$r=\heightlow{\mu}{\weight{\lambda}-\weight{\nu}}$, where
$\heightlow{\mu}{\weight{\lambda}-\weight{\nu}}$ measures the distance
from the lowest weight $\mu_*$ to $\weight{\lambda}-\weight{\nu}$ (see
Section~\ref{sec:exactwstates}). For these values of $r$, the
traced-out state is {\em exactly} given by a convex sum of
$\cW^r$-states.
\end{enumerate}

\begin{acknowledgments} \vspace{-0.2cm}
We thank Matthias Christandl, Ignacio Cirac, Tobias Osborne and Renato
Renner for helpful discussions. This work was supported by the EU project RESQ
(IST-2001-37559) and the European Commission through the FP6-FET
Integrated Project SCALA, CT-015714.  GM acknowledges support from the
project PROSECCO~(IST-2001-39227) of the IST-FET programme of the EC.
\end{acknowledgments}

\appendix

\subsection{Combinatorics and the upper
  bound in Corollary~\ref{cor:almostproduct}\label{app:combinatorial}}
In this appendix, we show that  
\begin{align}\label{eq:combinatoricsbnd}
\frac{\dimU{n-k}{d}}{\dimU{n}{d}}\sum_{i=r+1}^k\frac{\binom{k}{i}}{\binom{n}{i}}\binom{i+d-2}{i}\leq \frac{e^{3d}}{(d-2)!} \Bigl(\frac{k}{n-r}\Bigr)^{r+1} \Bigl(\frac{k(n-k)}{n-r}\Bigr)^{d-2}
\end{align}
if $d\leq \min\{k,n-k\}$, where $\dimU{n}{d}=\dim\Sym{\mathbb{C}^d}{n}=\binom{n+d-1}{n}$.

\begin{proof}
 We use the identity
\begin{align}
\sum_{i=r+1}^k\frac{\binom{k}{i}}{\binom{n}{i}}=\frac{k!(n-r)!}{(n-k+1)n!(k-r-1)!}
\end{align}
which  can be proven by observing that the identity holds trivially
for $r=k-1$ and by checking that  both sides satisfy the recursion relation
\[
a_r=a_{r-1}-\frac{\binom{k}{r}}{\binom{n}{r}}\ 
\]
as a function of $r$. We thus obtain
\begin{align}
\frac{\dimU{n-k}{d}}{\dimU{n}{d}}\sum_{i=r+1}^k\frac{\binom{k}{i}}{\binom{n}{i}}\binom{i+d-2}{i}\nonumber
\\
\leq
\frac{\dimU{n-k}{d}}{\dimU{n}{d}}\binom{k+d-2}{k}\sum_{i=r+1}^k\frac{\binom{k}{i}}{\binom{n}{i}}\label{eq:boundm}
\\
=\frac{1}{(d-2)!}\frac{(k+d-2)!(n-k+d-1)!(n-r)!}{(k-r-1)!(n-k+1)!(n+d-1)!}\label{eq:boundzero}\ .
\end{align}
We can further bound~\eqref{eq:boundzero}
as follows.
Note that
\[
\frac{(k+d-2)!}{(k-(r+1))!}\leq
k^{d+r-1}(1+\frac{d-2}{k})^{d+r-1}
\]
and thus with the inequality $(1+x)^y\leq e^{xy}$, 
\begin{align}
\frac{(k+d-2)!}{(k-(r+1))!}\leq
k^{d-1+r}e^{\frac{(d-1)^2}{k}}e^{\frac{(d-1)r}{k}}\ .\label{eq:boundone}
\end{align}
Similarly, we have
\begin{align}
\frac{(n-k+d-1)!}{(n-k+1)!}&\leq (n-k)^{d-2} e^{\frac{(d-1)^2}{n-k}}\label{eq:boundtwo}\\
\frac{(n-r)!}{(n+d-1)!} &\leq (n-r)^{-(d-1+r)}\label{eq:boundthree}\ .
\end{align}
Combining~\eqref{eq:boundzero}
with~\eqref{eq:boundone},~\eqref{eq:boundtwo}
and~\eqref{eq:boundthree}, we conclude that
\begin{align}\label{eq:mainalmostsymbound}
\frac{1}{(d-2)!}\bigl(\frac{k}{n-r}\bigr)^{d-1+r}(n-k)^{d-2}e^{\frac{(d-1)r}{k}+\frac{(d-1)^2}{k}+\frac{(d-1)^2}{n-k}}\ 
\end{align}
is an upper bound on the quantity~\eqref{eq:combinatoricsbnd}. The
claim then follows from this.
\end{proof}

It is interesting to note that for $d=2$, 
 inequality~\eqref{eq:boundm}
is tight, and  the  error~\eqref{eq:almostproductdelta} in
Corollary~\ref{cor:almostproduct} is  given by 
\[
2\Bigl(\frac{k!}{(k-r-1)!}\cdot\frac{(n-r)!}{(n+1)!}\Bigr)\qquad\textrm{for
  }d=2\ .
\]

\end{document}